\pdfoutput=1

\documentclass[12pt,a4paper]{article}

\usepackage{ifthen} 
\newboolean{pdflatex}
\setboolean{pdflatex}{true} 

\newboolean{articletitles}
\setboolean{articletitles}{true} 

\newboolean{uprightparticles}
\setboolean{uprightparticles}{false} 

\newboolean{inbibliography}
\setboolean{inbibliography}{false} 

\usepackage{microtype}
\usepackage{lineno}  
\usepackage{xspace} 
\usepackage{caption} 

\usepackage{graphicx}  
\usepackage{color}
\usepackage{colortbl}
\graphicspath{{./figs/}} 

\usepackage{amsmath} 
\usepackage{amssymb}
\usepackage{amsfonts}
\usepackage{upgreek} 

\newcommand*\patchAmsMathEnvironmentForLineno[1]{%
\expandafter\let\csname old#1\expandafter\endcsname\csname #1\endcsname
\expandafter\let\csname oldend#1\expandafter\endcsname\csname
end#1\endcsname
 \renewenvironment{#1}%
   {\linenomath\csname old#1\endcsname}%
   {\csname oldend#1\endcsname\endlinenomath}%
}
\newcommand*\patchBothAmsMathEnvironmentsForLineno[1]{%
  \patchAmsMathEnvironmentForLineno{#1}%
  \patchAmsMathEnvironmentForLineno{#1*}%
}
\AtBeginDocument{%
\patchBothAmsMathEnvironmentsForLineno{equation}%
\patchBothAmsMathEnvironmentsForLineno{align}%
\patchBothAmsMathEnvironmentsForLineno{flalign}%
\patchBothAmsMathEnvironmentsForLineno{alignat}%
\patchBothAmsMathEnvironmentsForLineno{gather}%
\patchBothAmsMathEnvironmentsForLineno{multline}%
\patchBothAmsMathEnvironmentsForLineno{eqnarray}%
}

\usepackage{hyperref}    
\usepackage[all]{hypcap} 

\def\lhcb {\mbox{LHCb}\xspace}

\def\babar  {\mbox{BaBar}\xspace}

\def\MagUp {\mbox{\em Mag\kern -0.05em Up}\xspace}

\ifthenelse{\boolean{uprightparticles}}%
{

 \def\Ppi         {\ensuremath{\uppi}\xspace}

 \def\PDelta      {\ensuremath{\Delta}\xspace}                 
 \def\PXi      {\ensuremath{\Xi}\xspace}                 
 \def\PLambda      {\ensuremath{\Lambda}\xspace}                 
 \def\PSigma      {\ensuremath{\Sigma}\xspace}                 
 \def\POmega      {\ensuremath{\Omega}\xspace}                 
 \def\PUpsilon      {\ensuremath{\Upsilon}\xspace}                 
 
 \def\PB      {\ensuremath{\mathrm{B}}\xspace}                 
                  
 \def\PD      {\ensuremath{\mathrm{D}}\xspace}

 \def\PK      {\ensuremath{\mathrm{K}}\xspace}

 \def\Pb      {\ensuremath{\mathrm{b}}\xspace}                 
 \def\Pc      {\ensuremath{\mathrm{c}}\xspace}

 \def\Pi      {\ensuremath{\mathrm{i}}\xspace}

 \def\Pp      {\ensuremath{\mathrm{p}}\xspace}

}
{

 \def\Ppi         {\ensuremath{\pi}\xspace}

 \mathchardef\PDelta="7101
 \mathchardef\PXi="7104
 \mathchardef\PLambda="7103
 \mathchardef\PSigma="7106
 \mathchardef\POmega="710A
 \mathchardef\PUpsilon="7107
                  
 \def\PB      {\ensuremath{B}\xspace}                 
                  
 \def\PD      {\ensuremath{D}\xspace}

 \def\PK      {\ensuremath{K}\xspace}

 \def\Pb      {\ensuremath{b}\xspace}                 
 \def\Pc      {\ensuremath{c}\xspace}

 \def\Pi      {\ensuremath{i}\xspace}

 \def\Pp      {\ensuremath{p}\xspace}

}

\makeatletter
\ifcase \@ptsize \relax
  \newcommand{\miniscule}{\@setfontsize\miniscule{4}{5}}
\or
  \newcommand{\miniscule}{\@setfontsize\miniscule{5}{6}}
\or
  \newcommand{\miniscule}{\@setfontsize\miniscule{5}{6}}
\fi
\makeatother

\DeclareRobustCommand{\optbar}[1]{\shortstack{{\miniscule (\rule[.5ex]{1.25em}{.18mm})}
  \\ [-.7ex] $#1$}}

\def\cquark    {{\ensuremath{\Pc}}\xspace}

\def\bquark    {{\ensuremath{\Pb}}\xspace}

\def\pion   {{\ensuremath{\Ppi}}\xspace}

\def\pip    {{\ensuremath{\pion^+}}\xspace}
\def\pim    {{\ensuremath{\pion^-}}\xspace}

\def\kaon    {{\ensuremath{\PK}}\xspace}
  \def\Kbar    {{\kern 0.2em\overline{\kern -0.2em \PK}{}}\xspace}

\def\KorKbar    {\kern 0.18em\optbar{\kern -0.18em K}{}\xspace}

\def\KS      {{\ensuremath{\kaon^0_{\rm\scriptscriptstyle S}}}\xspace}

  \def\Dbar    {{\kern 0.2em\overline{\kern -0.2em \PD}{}}\xspace}
\def\D       {{\ensuremath{\PD}}\xspace}

\def\DorDbar    {\kern 0.18em\optbar{\kern -0.18em D}{}\xspace}
\def\Dz      {{\ensuremath{\D^0}}\xspace}
\def\Dzb     {{\ensuremath{\Dbar{}^0}}\xspace}

\def\Dstarp  {{\ensuremath{\D^{*+}}}\xspace}
\def\Dstarm  {{\ensuremath{\D^{*-}}}\xspace}
\def\Dstarpm {{\ensuremath{\D^{*\pm}}}\xspace}

\def\Bbar    {{\ensuremath{\kern 0.18em\overline{\kern -0.18em \PB}{}}}\xspace}

\def\BorBbar    {\kern 0.18em\optbar{\kern -0.18em B}{}\xspace}

  \def\Y#1S{\ensuremath{\PUpsilon{(#1S)}}\xspace}

\def\proton      {{\ensuremath{\Pp}}\xspace}

\def\Lbar        {{\ensuremath{\kern 0.1em\overline{\kern -0.1em\PLambda}}}\xspace}
\def\LorLbar    {\kern 0.18em\optbar{\kern -0.18em \PLambda}{}\xspace}

\def\to                 {\ensuremath{\rightarrow}\xspace}

\def\CP                {{\ensuremath{C\!P}}\xspace}

\def\AT#1     {\ensuremath{A_{\mathrm{T}}^{#1}}\xspace}           

\def\C#1      {\ensuremath{\mathcal{C}_{#1}}\xspace}                       
\def\Cp#1     {\ensuremath{\mathcal{C}_{#1}^{'}}\xspace}                    
\def\Ceff#1   {\ensuremath{\mathcal{C}_{#1}^{\mathrm{(eff)}}}\xspace}        
\def\Cpeff#1  {\ensuremath{\mathcal{C}_{#1}^{'\mathrm{(eff)}}}\xspace}       
\def\Ope#1    {\ensuremath{\mathcal{O}_{#1}}\xspace}                       
\def\Opep#1   {\ensuremath{\mathcal{O}_{#1}^{'}}\xspace}                    

\newcommand{\ket}[1]{\ensuremath{|#1\rangle}}              

\newcommand{\tev}{\ifthenelse{\boolean{inbibliography}}{\ensuremath{~T\kern -0.05em eV}\xspace}{\ensuremath{\mathrm{\,Te\kern -0.1em V}}}\xspace}
\newcommand{\gev}{\ensuremath{\mathrm{\,Ge\kern -0.1em V}}\xspace}
\newcommand{\mev}{\ensuremath{\mathrm{\,Me\kern -0.1em V}}\xspace}
\newcommand{\kev}{\ensuremath{\mathrm{\,ke\kern -0.1em V}}\xspace}
\newcommand{\ev}{\ensuremath{\mathrm{\,e\kern -0.1em V}}\xspace}
\newcommand{\gevc}{\ensuremath{{\mathrm{\,Ge\kern -0.1em V\!/}c}}\xspace}
\newcommand{\mevc}{\ensuremath{{\mathrm{\,Me\kern -0.1em V\!/}c}}\xspace}
\newcommand{\gevcc}{\ensuremath{{\mathrm{\,Ge\kern -0.1em V\!/}c^2}}\xspace}
\newcommand{\gevgevcccc}{\ensuremath{{\mathrm{\,Ge\kern -0.1em V^2\!/}c^4}}\xspace}
\newcommand{\mevcc}{\ensuremath{{\mathrm{\,Me\kern -0.1em V\!/}c^2}}\xspace}

\def\mm   {\ensuremath{\rm \,mm}\xspace}

\def\mum  {\ensuremath{{\,\upmu\rm m}}\xspace}

\def\invfb   {\ensuremath{\mbox{\,fb}^{-1}}\xspace}

\def\ps   {\ensuremath{{\rm \,ps}}\xspace}
\def\fs   {\ensuremath{\rm \,fs}\xspace}

\def\invps{\ensuremath{{\rm \,ps^{-1}}}\xspace}

\def\gsim{{~\raise.15em\hbox{$>$}\kern-.85em
          \lower.35em\hbox{$\sim$}~}\xspace}
\def\lsim{{~\raise.15em\hbox{$<$}\kern-.85em
          \lower.35em\hbox{$\sim$}~}\xspace}

\def\ptot       {\mbox{$p$}\xspace}
\def\pt         {\mbox{$p_{\rm T}$}\xspace}

\def\evtgen     {\mbox{\textsc{EvtGen}}\xspace}

\def\geant      {\mbox{\textsc{Geant4}}\xspace}

\def\photos     {\mbox{\textsc{Photos}}\xspace}

\def\pythia     {\mbox{\textsc{Pythia}}\xspace}

\def\tell1  {TELL1\xspace}
\def\ukl1   {UKL1\xspace}

\newcommand{\eg}{\mbox{\itshape e.g.}\xspace}

\usepackage{cite} 
\usepackage{mciteplus}

\makeatletter
\g@addto@macro\bfseries{\boldmath}
\makeatother

\usepackage{longtable} 

\def\lnipchisq {\ensuremath{\ln\chi^2_{\mathrm{IP}}}\xspace}
\newcommand{\re}[2][()] {\ifthenelse{\equal{#1}{()}}{{\ensuremath{{\rm \, Re}}\left(#2\right)}}
                                                    {{\ensuremath{{\rm \, Re}}\left[#2\right]}}}
\newcommand{\im}[2][()] {\ifthenelse{\equal{#1}{()}}{{\ensuremath{{\rm \, Im}}\left(#2\right)}}
                                                    {{\ensuremath{{\rm \, Im}}\left[#2\right]}}}
\newcommand{\tmt}{\times 10^{-2}}
\newcommand{\cfit}{\texttt{cfit}}

\usepackage{float}

\begin{document}

\renewcommand{\thefootnote}{\fnsymbol{footnote}}
\setcounter{footnote}{1}


\begin{titlepage}
\pagenumbering{roman}

\vspace*{-1.5cm}
\centerline{\large EUROPEAN ORGANIZATION FOR NUCLEAR RESEARCH (CERN)}
\vspace*{1.5cm}
\noindent
\begin{tabular*}{\linewidth}{lc@{\extracolsep{\fill}}r@{\extracolsep{0pt}}}
\ifthenelse{\boolean{pdflatex}}
{\vspace*{-2.7cm}\mbox{\!\!\!\includegraphics[width=.14\textwidth]{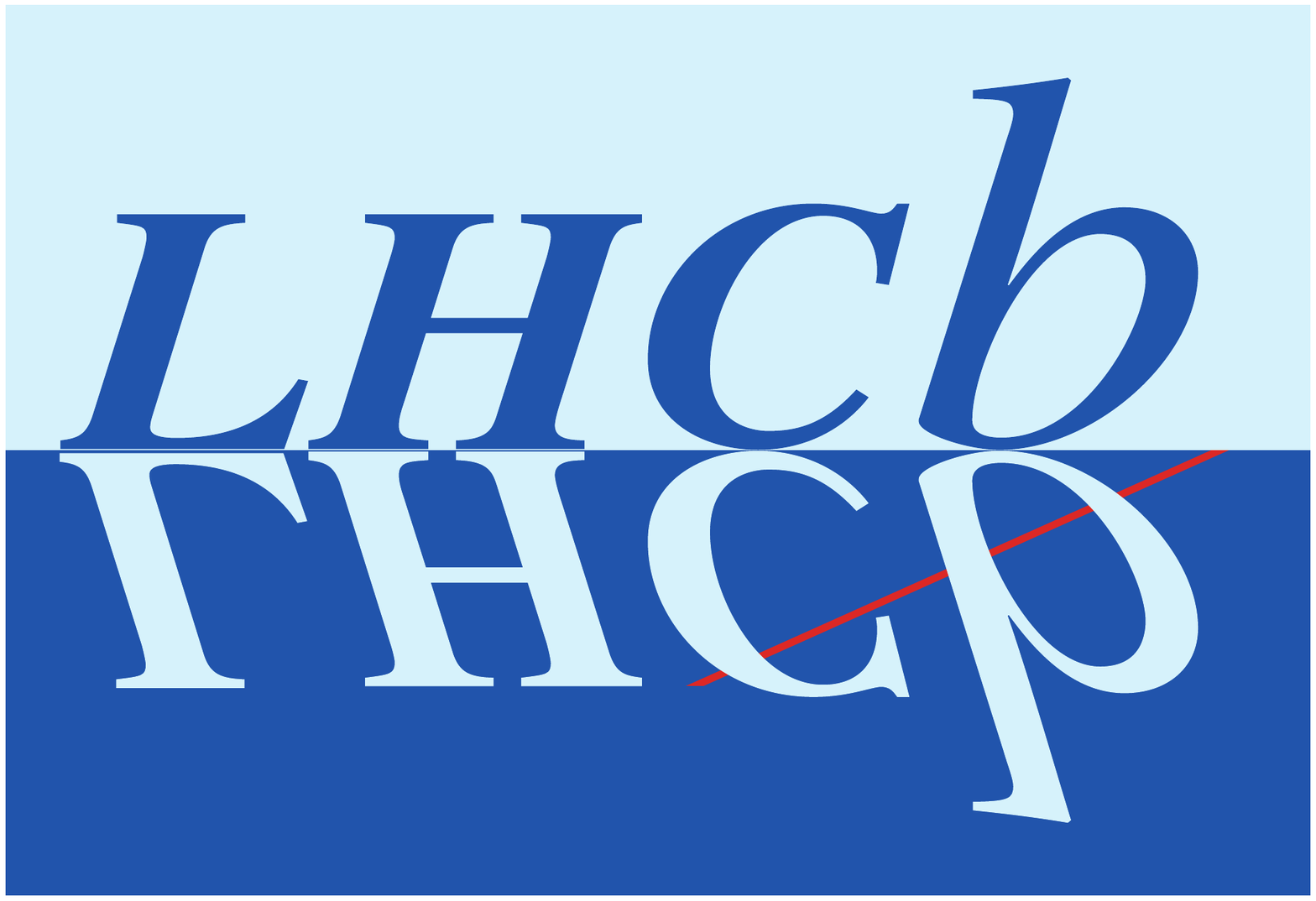}} & &}%
{\vspace*{-1.2cm}\mbox{\!\!\!\includegraphics[width=.12\textwidth]{lhcb-logo.eps}} & &}%
\\
 & & CERN-PH-EP-2015-249 \\  
 & & LHCb-PAPER-2015-042 \\  
 & & April 1, 2016 \\ 
 & & \\
\end{tabular*}

\vspace*{2.75cm}

{\bf\boldmath\huge
\begin{center}
  Model-independent measurement of
  mixing parameters
  in $\Dz \to \KS \pip \pim$ decays
\end{center}
}

\vspace*{1.5cm}

\begin{center}
The LHCb collaboration\footnote{Authors are listed at the end of this paper.}
\end{center}

\vspace{\fill}

\begin{abstract}
  \noindent
  The first model-independent measurement of the charm
  mixing parameters in the decay $\Dz \to \KS \pip \pim$
  is reported, using
  a sample of $\proton \proton$ collision data 
  recorded by the LHCb experiment,
  corresponding to an integrated luminosity of $1.0 \invfb$
  at a centre-of-mass energy of $7\tev$.
  The measured values are
  \begin{eqnarray*}
    x &=& \left( -0.86 \pm 0.53 \pm 0.17 \right) \tmt, \\
    y &=& \left( +0.03 \pm 0.46 \pm 0.13 \right) \tmt
          ,
  \end{eqnarray*}
  where the first uncertainties are statistical and
  include small contributions
  due to the external input for the strong phase measured
  by the CLEO collaboration,
  and the second uncertainties are systematic.
\end{abstract}

\vspace*{1.5cm}

\begin{center}
  Published in JHEP 03 (2016) 033.
\end{center}

\vspace{\fill}

{\footnotesize 
\centerline{\copyright~CERN on behalf of the \lhcb collaboration, licence \href{http://creativecommons.org/licenses/by/4.0/}{CC-BY-4.0}.}}
\vspace*{2mm}

\end{titlepage}


\newpage
\setcounter{page}{2}
\mbox{~}
%
%
%
%

\cleardoublepage


\renewcommand{\thefootnote}{\arabic{footnote}}
\setcounter{footnote}{0}



\pagestyle{plain} 
\setcounter{page}{1}
\pagenumbering{arabic}


%

\section{Introduction}
\label{sec:intro}

Mixing occurs in weakly decaying neutral mesons for which the flavour
eigenstates of the particle and antiparticle (\eg\ \Dz and \Dzb)
are not distinguished by any conserved quantum number.
It is characterised by the differences in mass, $\Delta M$,
and width, $\Delta \Gamma$, between
the mass eigenstates.
In the charm system these are usually expressed in a reduced form,
$x \equiv \Delta M / \Gamma$ and
$y \equiv \Delta \Gamma / (2 \Gamma)$,
where $\Gamma$ is the average of the two widths.

Mixing in charm has been observed with a significance above
five standard deviations in several independent
measurements~\cite{LHCb-PAPER-2012-038,Aaltonen:2013pja,LHCb-PAPER-2013-053,Ko:2014qvu}
and the constraints on $(x,y)$ are now rather precise~\cite{HFAG}.
However, most of the measurements are
sensitive to $(x^2 + y^2)$ or to $y$
(in the limit of negligible \CP violation),
leading to an ambiguity in the sign of $x$.
One approach to resolve this ambiguity is to exploit the
decay to the three-body, self-conjugate final state
$\Dz \to \KS \pip \pim$~\cite{Asner:2005sz,delAmoSanchez:2010xz,Peng:2014oda}.

The advantage of decays such as $\Dz \to \KS \pip \pim$ is that 
both Cabibbo-favoured (CF) and doubly Cabibbo-suppressed (DCS) components
are present in the same
final state. Therefore the strong phase differences between contributing
amplitudes---and hence between mixed and unmixed decays---can be measured with
an amplitude analysis~\cite{Muramatsu:2002jp,delAmoSanchez:2010xz,Peng:2014oda,Aaltonen:2012nd}
of the same data sample used to obtain the mixing parameters.
This is the approach that has been used to date.
A second method, proposed in Ref.~\cite{Bondar:2010qs}
and building upon a related approach for determining the
unitarity triangle angle $\gamma$\cite{Giri:2003ty},
uses measurements of the
average strong phase difference in regions of the phase space.
These can be obtained from an $e^+ e^-$ collider operating at
the $\psi(3770)$ resonance.
CLEO has made suitable measurements~\cite{Libby:2010nu}
and a similar study could be carried out with the larger
BESIII~\cite{Ablikim:2009aa} $\psi(3770)$ sample.
The advantage of this second method is that no amplitude analysis is needed:
the systematic uncertainty associated with the amplitude model
is replaced with the uncertainty on the strong phase measurements.
It has been estimated that with BESIII data this external uncertainty should be
smaller than the statistical uncertainty for $\Dz \to \KS \pip \pim$
yields of up to 10--20 million~\cite{Thomas:2012qf},
far larger than those available today.
This paper describes the first measurement of $x$ and $y$
with this novel method,
using promptly produced charm mesons in the decay chain
$\Dstarp \to \Dz \pip, \Dz \to \KS \pip \pim, \KS \to \pip \pim$
(charge conjugate processes are included implicitly
unless otherwise noted).
A sample of $\proton \proton$ collision data 
recorded by the LHCb experiment in 2011 is used,
corresponding to an integrated luminosity of $1.0 \invfb$
at a centre-of-mass energy of $7\tev$.

\section{Formalism}
\label{sec:formalism}

\newcommand{\Adz} {\mathcal{A}_{\Dz} }
\newcommand{\Adzb}{\mathcal{A}_{\Dzb}}

The formalism for the method has been presented
previously~\cite{Bondar:2010qs,Thomas:2012qf,Libby:2010nu},
but is summarised here for clarity.
The flavour eigenstates,
$|\Dz\rangle$ and $|\Dzb\rangle$,
are related to the mass eigenstates,
$|D_1\rangle$ and $|D_2\rangle$,
via
\begin{align}
  |D_1\rangle &= p|\Dz\rangle - q|\Dzb\rangle, \\
  |D_2\rangle &= p|\Dz\rangle + q|\Dzb\rangle,
\end{align}
where $|p|^2+|q|^2=1$.
In the limit of \CP conservation, $|p/q|=1$. There is one free phase that is fixed
by stipulating that in the limit of no indirect \CP violation,
  $q/p = +1$ and $\ket{D_1}$ is the \CP-odd eigenstate.
The sign convention adopted for the mixing parameters is
\begin{eqnarray}
  x &=& ( M_2 - M_1 ) / \Gamma , \\
  y &=& ( \Gamma_2 - \Gamma_1 ) / (2\Gamma)
  .
\end{eqnarray}

For a state that is initially pure \Dz at $t=0$, let the state
at some later time $t$ be $|\Dz(t)\rangle$.
Likewise, let the time evolution of \Dzb be $|\Dzb(t)\rangle$.
These may be evaluated as
\begin{eqnarray}
  |\Dz(t) \rangle &=& \phantom{\frac{p}{q}}\,g_+(t)|\Dz\rangle + \frac{q}{p}          \,g_-(t)|\Dzb\rangle , \label{eq:formalism:timeEvA} \\
  |\Dzb(t)\rangle &=& \frac{p}{q}          \,g_-(t)|\Dz\rangle + \phantom{\frac{q}{p}}\,g_+(t)|\Dzb\rangle \label{eq:formalism:timeEvB}
  ,
\end{eqnarray}
where
\begin{equation}
  g_\pm(t) \equiv \frac{e^{-i(M_2 - i\Gamma_2/2)t} \pm e^{-i(M_1 - i\Gamma_1/2)t}}{2}
  .
  \label{eq:formalism:defineG}
\end{equation}

The phase space for the three-body decay of a \Dz or \Dzb meson to
$\KS \pip \pim$ is conventionally represented as a Dalitz
plot and can be described by two variables,
$m^2_{12} = m^2(\KS \pip)$ and $m^2_{13} = m^2(\KS \pim)$.
Let the amplitude for a \Dz decay to a point
$(m^2_{12},m^2_{13})$ in the phase space be
$  \mathcal{A}_{\Dz}(m_{12}^2,m_{13}^2)   $.
Neglecting direct \CP violation, the amplitudes for \Dz and \Dzb
are related by the exchange $m^2_{12} \leftrightarrow m^2_{13}$,
\begin{equation}
  \mathcal{A}_{\Dzb}(m_{12}^2,m_{13}^2) = \mathcal{A}_{\Dz}(m_{13}^2,m_{12}^2)
  .
  \label{eq:formalism:defineADzb}
\end{equation}
In the expressions that follow, the explicit dependence of the amplitude terms
$\Adz$ and $\Adzb$ on $m^2_{12}$ and $m^2_{13}$ is omitted.
The amplitude $A_{\Dz} (m^2_{12}, m^2_{13}, t)$
for a state that was initially \Dz to decay at some later
time $t$ to a point $(m^2_{12}, m^2_{13})$ in the phase space is
\begin{equation}
  A_{\Dz} (m^2_{12}, m^2_{13}, t) 
  = \mathcal{A}_{\Dz} \, g_+(t) + \frac{q}{p} \mathcal{A}_{\Dzb} \, g_-(t)
  .
\end{equation}
Similarly,
\begin{equation}
  A_{\Dzb} (m^2_{12}, m^2_{13}, t) 
  = \mathcal{A}_{\Dzb} \, g_+(t) + \frac{p}{q} \mathcal{A}_{\Dz} \, g_-(t)
  .
\end{equation}
The probability density ${\cal {P}}_{\Dz}(m_{12}^2, m_{13}^2, t)$
is given by the modulus squared of the amplitude multiplied
by a normalisation factor of $\Gamma$,
\begin{equation}
  {\cal {P}}_{\Dz}(m_{12}^2, m_{13}^2, t) = \Gamma \left| A_\Dz(m_{12}^2, m_{13}^2, t) \right|^2
  ,
\end{equation}
with ${\cal {P}}_{\Dzb}$ defined similarly in terms of $A_\Dzb$.
Performing a Taylor expansion and
neglecting terms of order $x^2$, $xy$, and $y^2$, these evaluate to
\begin{eqnarray}
  {\cal {P}}_{\Dz}(m_{12}^2, m_{13}^2, t) &=&
  \Gamma e^{-\Gamma t}
  \left[
    \left| \Adz \right|^2 - \Gamma t \, \re{\frac{q}{p} \Adz^\star \Adzb ( y + i\,x )}
  \right]
  ,
            \\
  {\cal {P}}_{\Dzb}(m_{12}^2, m_{13}^2, t) &=&
  \Gamma e^{-\Gamma t}
  \left[
    \left| \Adzb \right|^2 - \Gamma t \, \re{\frac{p}{q} \Adz \Adzb^\star ( y + i\,x )}
  \right]
  .
\end{eqnarray}
Neglecting \CP violation for the purposes of the mixing measurement,
$q/p=1$ and hence
\begin{eqnarray}
  {\cal {P}}_{\Dz}(m_{12}^2, m_{13}^2, t) &=&
  \Gamma e^{-\Gamma t}
  \left[
    \left| \Adz \right|^2 - \Gamma t \, \re{\Adz^\star \Adzb ( y + i\,x )}
  \right] ,
  \\
  {\cal {P}}_{\Dzb}(m_{12}^2, m_{13}^2, t) &=&
  \Gamma e^{-\Gamma t}
  \left[
    \left| \Adzb \right|^2 - \Gamma t \, \re{\Adzb \Adz^\star ( y + i\,x )}
  \right]
  .
\end{eqnarray}

These densities may be integrated over regions of the phase space.
Various binning schemes are possible; this analysis uses the
one referred to as the ``equal $\Delta \delta_D$ \babar 2008'' binning
in Ref.~\cite{Libby:2010nu}, in which the strong phase
variation within each bin of the phase space is minimised.
This has the advantage of reducing the sensitivity to detector effects
  such as variation in efficiency across the phase space.
In this scheme there are 16 bins, with bins 1 to 8 in the
region of the phase space $m_{12}^2 > m_{13}^2$ and bins
$-1$ to $-8$ in the region $m_{12}^2 < m_{13}^2$. The bins are symmetric
about the leading diagonal, with bin $i$ mapped to bin $-i$
by the transformation $(m_{12}^2, m_{13}^2) \to (m_{13}^2, m_{12}^2)$.
The quantities $T_i$ and $X_i$ are defined by the integrals
\begin{align}
  T_i &\equiv                             \int_i \left| \Adz \right|^2 \mathrm{d}m^2_{12} \, \mathrm{d}m^2_{13} , \label{eq:formalism:defineTi} \\
  X_i &\equiv \frac{1}{\sqrt{T_i T_{-i}}} \int_i \Adz^\star \, \Adzb \, \mathrm{d}m^2_{12} \, \mathrm{d}m^2_{13} \, 
  ,
\end{align}
and the $X_i$ may in turn be
expressed in terms of real quantities $c_i$ and $s_i$ as
\begin{align}
  c_i &\equiv   \re{X_i}, \\
  s_i &\equiv - \im{X_i}.
\end{align}
Given the symmetric binning, Eq.~\ref{eq:formalism:defineADzb} implies that
$X_{-i} = X_i^\star$,
and thus $c_{-i} = c_i$ and $s_{-i} = -s_i$.

With these definitions, the integrated probability densities are
\begin{align}
  {\cal {P}}_{\Dz}(i; t)
  &= \int_i {\cal {P}}_{\Dz}(m_{12}^2, m_{13}^2, t) \, \mathrm{d}m^2_{12} \, \mathrm{d}m^2_{13} \nonumber \\
  &= \Gamma \, e^{-\Gamma t} \left[ T_i - \Gamma t \sqrt{T_i T_{-i}} \left\{ y c_i + x s_i \right\} \right]
    ,
  \label{eq:formalism:timeDependenceForDz}
\end{align}
and
\begin{equation}
  {\cal {P}}_{\Dzb}(i; t)
  = \Gamma \, e^{-\Gamma t} \left[ T_{-i} - \Gamma t \sqrt{T_i T_{-i}} \left\{ y c_i - x s_i \right\} \right]
  \label{eq:formalism:timeDependenceForDzb}
  .
\end{equation}
These distributions are used to obtain the mixing
parameters $x$ and $y$.
The values of $T_i$, $c_i$, and $s_i$ measured by the CLEO collaboration
are given in Tables~VII and XVI of Ref.~\cite{Libby:2010nu}.\footnote{
  Note that the captions for Tables~VII and VIII
  were exchanged in Ref.~\cite{Libby:2010nu},
  and that the supplementary material defining the binning
  contains an off-by-one error in the bin indices.
}

\section{Detector, selection and simulation}
\label{sec:Detector}

The \lhcb detector~\cite{Alves:2008zz,LHCb-DP-2014-002} is a single-arm forward
spectrometer covering the \mbox{pseudorapidity} range $2<\eta <5$,
designed for the study of particles containing \bquark or \cquark
quarks. The detector includes a high-precision tracking system
consisting of a silicon-strip vertex detector surrounding the $pp$
interaction region, a large-area silicon-strip detector located
upstream of a dipole magnet with a bending power of about
$4{\rm\,Tm}$, and three stations of silicon-strip detectors and straw
drift tubes placed downstream of the magnet.
The tracking system provides a measurement of momentum, \ptot, of charged particles with
a relative uncertainty that varies from 0.5\% at low momentum to 1.0\% at 200\gevc.
The minimum distance of a track to a primary vertex, the impact parameter~(IP), is measured with a resolution of $(15+29/\pt)\mum$,
where \pt is the component of the momentum transverse to the beam, in\,\gevc.
Different types of charged hadrons are distinguished using information
from two ring-imaging Cherenkov detectors. 
Photons, electrons and hadrons are identified by a calorimeter system consisting of
scintillating-pad and preshower detectors, an electromagnetic
calorimeter and a hadronic calorimeter. Muons are identified by a
system composed of alternating layers of iron and multiwire
proportional chambers.

The online event selection is performed by a trigger~\cite{LHCb-DP-2012-004}, 
which consists of a hardware stage, based on information from the calorimeter and muon
systems, followed by a software stage, which applies a full event
reconstruction.
At the hardware trigger stage, events are required to have a muon with high \pt or a
  hadron, photon or electron with high transverse energy in the calorimeters.
In the subsequent software trigger, 
  pairs of oppositely charged tracks are combined to form \KS candidates,
  and those are in turn combined with a second pair of oppositely charged tracks
  to form \Dz candidates. 
  For the 2011 dataset, the trigger requires that all four tracks be
  reconstructed in the vertex detector, reducing the \KS efficiency
  significantly.
  Both the \KS and \Dz candidate vertices are required to be displaced from 
  any primary $pp$ interaction vertex~(PV) in the event,
  and additional geometrical and kinematic critera are applied 
  to suppress background and ensure consistency with a $\Dz \to \KS \pip \pim$ decay.
  These include a requirement that at least one of the four tracks has an
  impact parameter larger than $100\mum$ with respect to any PV.

After offline processing, additional selection criteria are applied to further
suppress background. 
These include particle identification requirements on the \Dz daughter tracks,
as well as requirements
  that the track and vertex fits be of good quality,
  that the \KS vertex be at least $10\mm$ downstream of the \Dz vertex,
  that the \KS candidate mass lie within $\pm 11.4\mevcc$ of the known value~\cite{PDG2014},
  that the \Dz candidate mass $m_D$ lie within $\pm 85\mevcc$ of the known value~\cite{PDG2014},
  and that the reconstructed \Dz decay time $t_D$ lie within $0.3 < t_D < 5$\ps.
The \Dz candidate is also required to have no more than two turning points in its
  decay time acceptance function (see Sec.~\ref{sec:fit:time:swimming}).
It is then combined with a fifth pion track, referred to as the
  soft pion, to form a \Dstarp candidate. 
  Both the soft pion and \Dz candidate are constrained to originate from the same PV.
  Good vertex fit quality is required, and
  particle identification requirements are applied to the soft pion.
  The mass difference $\Delta m = m_{\Dstarp} - m_D$ is required to lie within the range
  $m_{\pi} < \Delta m < (m_{\pi} + 15\mevcc)$, where $m_{\Dstarp}$ is the mass of the \Dstarp candidate
  and $m_{\pi}$ is the charged pion mass.
  If there is more than one distinct $\Dz \to \KS \pip \pim$ candidate
  then one is chosen at random and the rest are discarded. If, after this, there are multiple
  \Dstarpm candidates then the one with the best vertex fit quality is retained and the rest are discarded.

Simulated events are used for cross-checks.
In the simulation, $pp$ collisions are generated using
\pythia~6~\cite{Sjostrand:2006za}
with a specific \lhcb
configuration~\cite{LHCb-PROC-2010-056}.  Decays of hadronic particles
are described by \evtgen~\cite{Lange:2001uf}, in which final-state
radiation is generated using \photos~\cite{Golonka:2005pn}. The
interaction of the generated particles with the detector, and its response,
are implemented using the \geant
toolkit~\cite{Allison:2006ve, *Agostinelli:2002hh} as described in
Ref.~\cite{LHCb-PROC-2011-006}.

\section{Fits}
\label{sec:fit}

\subsection{Overview}
\label{sec:fit:intro}

The mixing parameters $x$ and $y$ are determined by a sequence of
fits to the distributions of
the variables
  $(m_D, \, \Delta m)$ and $(t_D, \, \lnipchisq)$,
  initially in the whole phase space and later in individual regions.
The impact parameter $\chi^2$, $\chi^2_{\mathrm{IP}}$,
  is defined as the difference in the vertex fit $\chi^2$ of the associated
  PV with and without the \Dz candidate.
It is used to separate prompt charm
  that originates at the PV
  from secondary charm produced at a displaced vertex.
The dominant source of secondary charm is from decays of
  $b$-hadrons.
Two other variables are also used to describe
  the per-event decay time acceptance function,
  introduced in Sec.~\ref{sec:fit:time:swimming}.
Unless otherwise specified, all data passing the selection described
in Sec.~\ref{sec:Detector} are used. Where reference is made to
a narrow signal window in $m_D$ or $\Delta m$, this corresponds to
a stricter requirement:
$\pm 20\mevcc$ around the known \Dz mass, or
$144.2 < \Delta m < 146.4\mevcc$.
The mass sidebands are defined as
$1785 < m_D < 1810\mevcc$ and $1920 < m_D < 1945\mevcc$.

First, an extended maximum likelihood fit to the $m_D$ distribution
  of all selected \Dstarp candidates is performed
  to determine
  the amounts of \Dz signal and combinatorial background in
  the narrow $m_D$ signal window (Sec.~\ref{sec:fit:mass}).
Second is a maximum likelihood fit to the $(t_D, \lnipchisq)$ distribution
  of those candidates in the narrow $m_D$ signal window,
  using the mass sidebands to estimate the background distributions (Sec.~\ref{sec:fit:time:lifetime}).
  This fit uses the yields determined in the first fit,
  and serves to determine the \lnipchisq shapes for
  prompt and secondary charm. It is not sensitive to mixing.
Third is a set of 32 extended maximum likelihood fits, each to the
  $(m_D, \Delta m)$ distribution in a particular phase space
  bin, with the \Dstarp and \Dstarm samples fitted separately (Sec.~\ref{sec:fit:massAndDeltaMass}).
  Each fit provides measurements of the amounts of signal and background in
  the narrow $(m_D, \Delta m)$ window for the corresponding bin.
Fourth is a simultaneous maximum likelihood fit to the $(t_D, \lnipchisq)$
  distributions of candidates for the 32 subsamples (Sec.~\ref{sec:fit:time:mixing}). 
  Signal candidates are required to lie
  in the narrow $m_D$ and $\Delta m$ signal windows, with the
  mass sidebands used to constrain the combinatorial background.
  This fit uses the $\lnipchisq$
  shapes from the second fit and the yield estimates from the third
  fit, and produces measurements of
  the mixing parameters $x$ and~$y$.

Only the fit procedure and results are discussed in this section.
Cross-checks and systematic uncertainties are discussed in
Sec.~\ref{sec:sys}.
All aspects of the selection and fit procedure were finalised
before any measurements of $x$ and $y$ were made.
Unless otherwise stated, all parameters introduced are left
free in the fits.

\subsection{Fit to $m_D$}
\label{sec:fit:mass}

The probability density functions (PDFs) used to model the $m_D$
distributions
are expressed in terms of
  exponential,
  Gaussian ($G$),
  bifurcated Gaussian ($B$), and
  Crystal Ball ($C$)~\cite{Skwarnicki:1986xj} functions.
Only two components are needed:
\Dz signal and combinatorial background.
The PDF for \Dz signal~(sig) is the sum of a Gaussian, a bifurcated Gaussian,
and a Crystal Ball function,
\begin{equation}
  f_1(m_D|\mathrm{sig}) = \eta_1 \, G(m_D ; \mu_D, \sigma_1) + \eta_2 \, B( m_D ; \mu_D, \sigma_L, \sigma_R) 
                + (1 - \eta_1 - \eta_2) \, C( m_D ; \mu_D, \sigma_2, \alpha, n)
,
\end{equation}
where the order of the Crystal Ball function, $n$, is fixed to three.
The PDF for the combinatorial background, $f_1(m_D|\mathrm{cmb})$, is an exponential function.
The total PDF is then
\begin{equation}
  f_1(m_D) =  P_1(\mathrm{sig}) \, f_1(m_D|\mathrm{sig}) + P_1(\mathrm{cmb}) \, f_1(m_D|\mathrm{cmb}) 
  ,
\end{equation}
where $P_1(\mathrm{sig})$ and $P_1(\mathrm{cmb})$ describe the fractions of
signal and background in the data sample used for the first fit, and sum to unity.

\begin{figure}
\centering
\includegraphics[width = 0.49\linewidth]{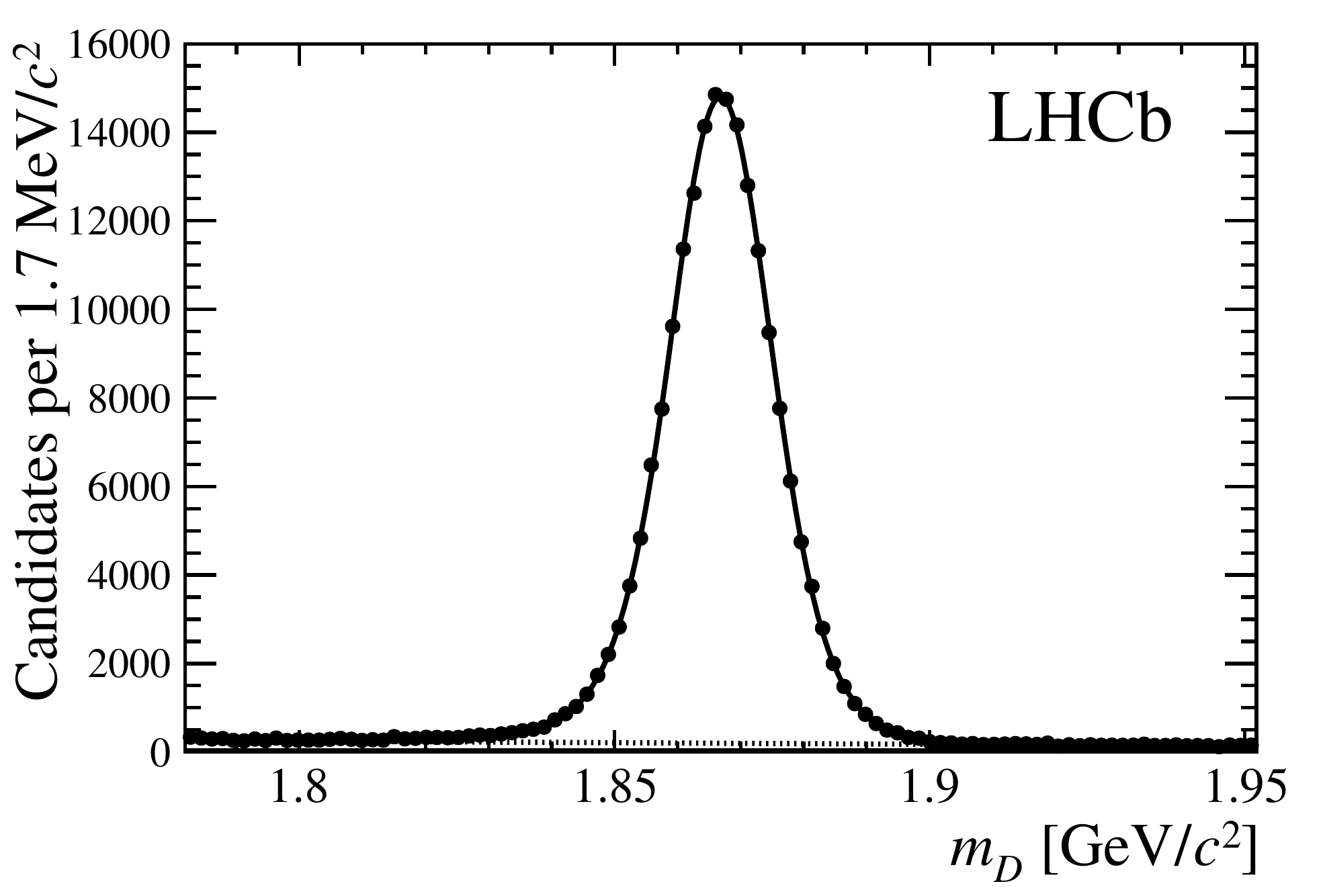}
\includegraphics[width = 0.49\linewidth]{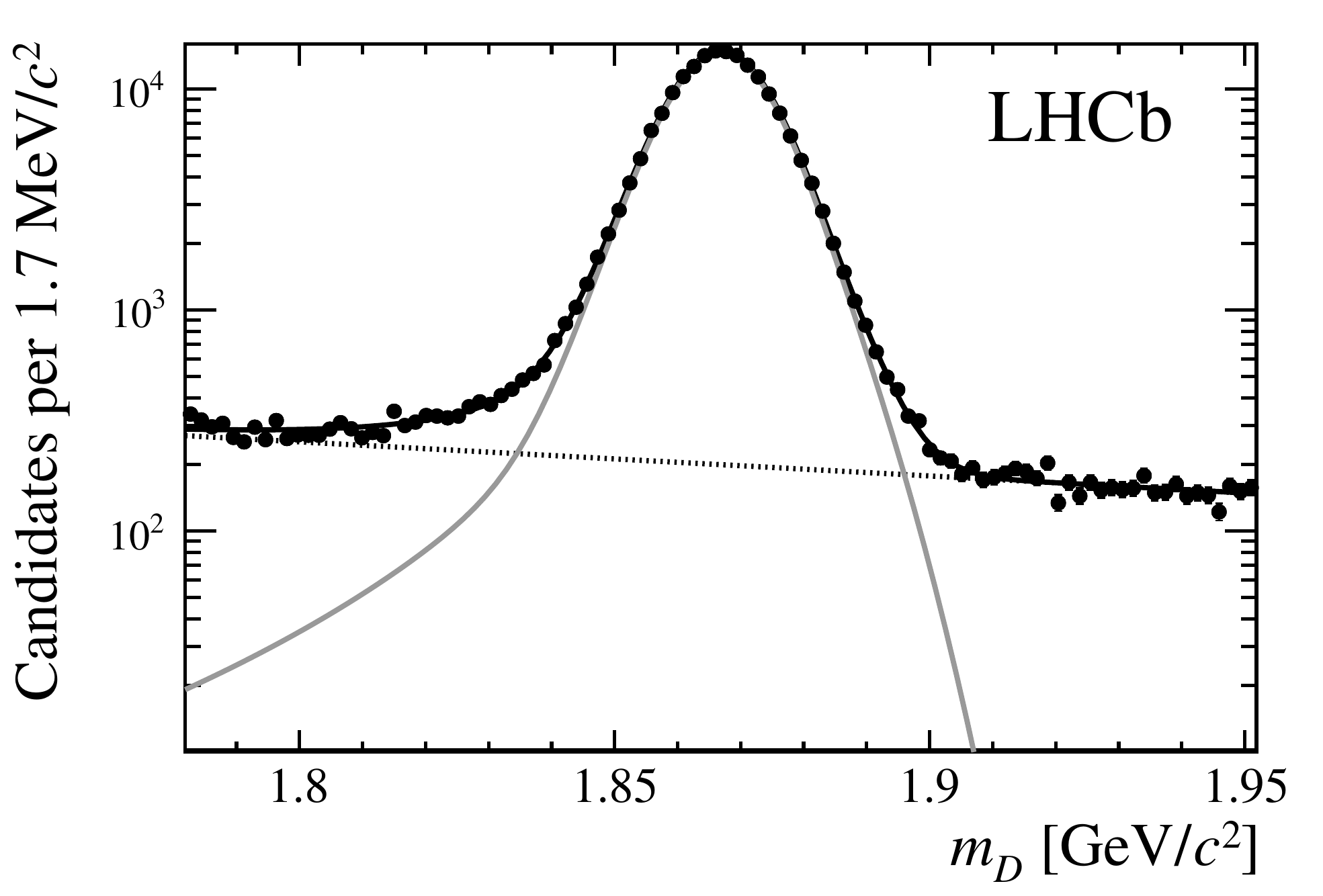}
\caption{
  Fitted $m_D$ distribution.
  Both plots show the same data sample with (left) linear and
  (right) logarithmic vertical scales.
  The curves show the results of the first fit, described in Sec.~\ref{sec:fit:mass}:
  the total (solid black),
  the background component (dotted),
  and the signal component (grey, right only).
}
\label{fig:fit:mass:fit1}
\end{figure}

The results of the first fit are shown in 
Fig.~\ref{fig:fit:mass:fit1}.
The fit yields 178k
signal events within the narrow $m_D$ signal window,
and the purity within this window is $(97.4 \pm 0.3) \%$.

\subsection{Time acceptance correction}
\label{sec:fit:time:swimming}

The probability for a \Dstarp signal decay to be
successfully triggered, reconstructed, and selected depends upon the decay time
of its \Dz daughter.
The time-dependent fits must, therefore, take account of the nonuniform decay time acceptance.
A data-driven method referred to as swimming~\cite{bib:swimming} is used.
This approach follows that used in previous LHCb measurements of the mixing and
indirect \CP violation parameters, $y_{\CP}$ and $A_{\Gamma}$, in \Dz
decays~\cite{LHCb-PAPER-2011-032,LHCb-PAPER-2013-054},
and at previous experiments~\cite{Bailey:1985zz,Adam:1995mb,Rademacker:2005ay,Aaltonen:2010ta}.

The principle of the method is that the decay time acceptance is determined
by selection criteria that can be reproduced later. The criteria for this analysis are
given in Sec.~\ref{sec:Detector}.
(In practice they are applied to the measured rather than the true decay time;
the resolution is neglected and considered as a systematic effect
(Sec.~\ref{sec:sys}).)
Those criteria can be tested again after modifying the candidate---specifically,
with a different decay time. By repeatedly testing the criteria for many decay
time values spanning the allowed range, the acceptance function
for an individual candidate may be determined empirically.
Aside from a correction factor discussed later in this section,
the value of this
function is 1 for those decay times at which all of the criteria are fulfilled,
and 0 at all other times.
Since candidates with $t_D < 0.3\ps$ are rejected, the
acceptance function is zero below that point. It must also be zero at very
large decay times, both because of the upper bound on $t_D$ and because of the
finite length of the vertex detector. Therefore, the acceptance function will
take the form of a top-hat function
  [$\Theta(t_D - t_0) - \Theta(t_D - t_1)$], where $\Theta$ is the Heaviside
   function and $t_1 > t_0$,
or will be the sum of several nonoverlapping top-hat functions.
The decay times at which the acceptance changes between 0 and 1 are referred
to as the turning points.

For approximately 90\% of selected candidates, the acceptance is a single top-hat
with exactly two turning points. The remaining candidates have a more complicated
acceptance function, typically due to the presence of a second $\proton \proton$
primary vertex nearby. As in the previous analysis using this
technique~\cite{LHCb-PAPER-2013-054},
candidates with more than two turning points are rejected.
This enables a more robust description of the turning point variable
distributions (see below) and suppresses events in which the primary vertex
association is ambiguous.

The implementation of the decay time acceptance calculation is simplified
by a number of assumptions.
  First, the hardware triggers do not depend
    on the \Dz decay time and can therefore be ignored when evaluating the
    acceptance function.
  Second, the decay time acceptance depends only on
    the \Dz reconstruction and selection: it is not affected by the soft pion and
    \Dstarp requirements.
  Third, the full vertex detector pattern recognition is not re-run when
    changing the \Dz decay time; instead, the changes to the decay geometry
    are made analytically. Requirements on the number of hits on a track
    in the vertex detector subsystem are approximated as requirements
    that the modified trajectory pass through a corresponding number of
    subdetector modules.
  Finally, an additional correction factor $\varepsilon(t_D)$
    is applied to the acceptance function
    to model the effect of a track quality cut in the reconstruction, which
    reduces the efficiency for tracks produced further from the beam axis. The correction
    is derived from samples of simulated events and is parameterised
    as a polynomial function.

For an individual event, the acceptance
function can be written as
\begin{equation}
  a(t_D ; t_0, \Delta t) = \left[ \Theta(t_D - t_0) - \Theta(t_D - t_0 - \Delta t) \right] \varepsilon(t_D)
  ,
  \label{eq:fit:time:swimming:perEventAcceptanceFunction}
\end{equation}
where $t_0$ is the first turning point (TP) and $\Delta t$ is the difference
between the two turning points.
Although the acceptance function is determined for each event independently,
models of the distribution $f_{\mathrm{TP}} (t_0, \Delta t)$
of the turning point variables $t_0$ and $\Delta t$
are required for the decay time fits.
The distribution is assumed to factorise,
\begin{equation}
  f_{\mathrm{TP}}(t_0, \Delta t) = f_{\mathrm{TP,0}}(t_0) \, f_{\mathrm{TP,\Delta}} (\Delta t)
  .
\end{equation}
Nonparametric functions are used to model the turning point PDFs.
The distribution $f_{\mathrm{TP,0}}(t_0)$ is modelled as a histogram PDF
with 100 bins spanning the range $0$--$3\ps$ and the
distribution $f_{\mathrm{TP,\Delta}} (\Delta t)$
is modelled as a one-dimensional Gaussian kernel PDF~\cite{bib:DensityEstimation}.
The same method is used for all components,
and is based on data in the mass sidebands
for combinatorial background.
Candidates in the narrow mass signal window are used for 
prompt and secondary \Dz mesons,
both of which are assumed to have the same
turning point distribution in the baseline fit.

\subsection{Separation of prompt and secondary candidates}
\label{sec:fit:time:lifetime}

The second fit
is used to determine the relative proportions of prompt
and secondary \Dz signal, and to model
their \lnipchisq distributions.
It also serves as an important cross-check since it allows
the mean \Dz lifetime to be computed in the $\Dz \to \KS \pip \pim$ sample.
No distinction is made in the fit between \Dz and \Dzb candidates,
nor between different regions of the phase space, so by design it is
insensitive to mixing. While a dominance of \CP-odd or of \CP-even
components in the final state could in principle shift the mean lifetime by up to
$\pm \, y \, \Gamma \approx 2.5\fs$, the net \CP has recently been
shown to be almost zero~\cite{Gershon:2015xra} so that the effective lifetime is close to $\tau_D$.
Similarly, previous amplitude analyses all found
that the decay is dominated by flavour-specific processes,
with total fit fractions of about
70\%~\cite{Muramatsu:2002jp,delAmoSanchez:2010xz,Peng:2014oda,Aaltonen:2012nd},
implying that the maximum scale of the effect is below the sensitivity of this analysis.

In this fit, the underlying decay time distribution for the prompt~(prm) \Dz signal
is taken to be an exponential function for $t_D > 0$ with characteristic time $\tau_D$.
For a particular event, the expected $t_D$ distribution is
this exponential multiplied by the per-event acceptance function
given in Eq.~\ref{eq:fit:time:swimming:perEventAcceptanceFunction},
\begin{equation}
  f_2(t_D | t_0, \Delta t ; \mathrm{prm}) = n \, a(t_D ; t_0, \Delta t) \, e^{- t_D / \tau_D}
  ,
  \label{eq:fit:time:lifetime:promptTimeDistributionRaw}
\end{equation}
where $n$ is a normalisation factor
and the decay time resolution has been neglected.
Note that the expression in Eq.~\ref{eq:fit:time:lifetime:promptTimeDistributionRaw}
depends explicitly on the turning point variables $t_0$ and $\Delta t$. 
To separate out this dependence, the models for the turning point distributions
given in Sec.~\ref{sec:fit:time:swimming} are used.
The PDF for prompt charm may then be written as
\begin{equation}
\begin{split}
  f_2(t_0, \Delta t, t_D, \lnipchisq | \mathrm{prm}) = &
    f_2(\lnipchisq | t_D ; \mathrm{prm}) \,
    f_2(t_D | t_0, \Delta t ; \mathrm{prm}) \\
    & \times f_{\mathrm{TP,0}} (t_0 | D) \,
    f_{\mathrm{TP,\Delta}} (\Delta t | D)
    ,
\end{split}
\end{equation}
where
$D$ denotes PDFs used for both prompt and secondary \Dz, and
$f_2(\lnipchisq | t_D ; \mathrm{prm})$ is a parameterisation of the
\lnipchisq distribution for a given decay time, taking the form
\begin{equation}
  f_2(\lnipchisq | t_D ; \mathrm{prm}) 
    = \eta     \, G(\lnipchisq ; \mu_{\mathrm{p}}(t_D), \sigma_1)
    + (1-\eta) \, B(\lnipchisq ; \mu_{\mathrm{p}}(t_D), \sigma_L, \sigma_R)
    ,
    \label{eq:fit:time:lifetime:promptlnipchisq}
\end{equation}
where $\mu_{\mathrm{p}}(t_D)$,
the most probable value of \lnipchisq,
is a linear function.

A similar approach is used for the secondary~(sec) \Dz signal, except that the
underlying decay time distribution is taken to be the convolution of two
exponential functions restricted to $t_D > 0$
and with characteristic times $\tau_1$ and $\tau_2$.
Since
$\left[ \Theta(t_D) \, e^{-t_D/\tau_1} \right] \otimes \left[ \Theta(t_D) \, e^{-t_D/\tau_2} \right]$
may be rewritten as
$(e^{-t_D/\tau_2} - e^{-t_D/\tau_1})$
with an appropriate normalisation factor,
the expression remains analytically integrable and takes the form
\begin{equation}
  f_2(t_D | t_0, \Delta t ; \mathrm{sec}) = n \, a(t_D ; t_0, \Delta t) \,
    \left( 
      e^{- t_D / \tau_2} - e^{- t_D / \tau_1}
    \right)
  ,
\end{equation}
where $n$ is again a normalisation factor.
The \lnipchisq distribution also differs from that used for prompt charm,
\begin{equation}
  f_2(\lnipchisq | t_D ; \mathrm{sec}) 
    = \eta     \, G(\lnipchisq ; \mu_{\mathrm{s}}(t_D), \alpha \, \sigma_1)
    + (1-\eta) \, B(\lnipchisq ; \mu_{\mathrm{s}}(t_D), \alpha \, (\sigma_L + \beta t_D), \alpha \, \sigma_R)
    .
\end{equation}
Compared to Eq.~\ref{eq:fit:time:lifetime:promptlnipchisq},
the width of the peak is multiplied by $\alpha$,
with the lower tail of the bifurcated Gaussian having a further, time-dependent broadening.
In addition, the decay time at which the function is maximised,
$\mu_{\mathrm{s}}(t_D)$, is taken empirically to evolve as
\begin{equation}
  \mu_{\mathrm{s}}(t_D) = \mu_{\mathrm{s} 0} + B (1 - e^{C t_D} )
  .
\end{equation}
Using the models for the turning point distributions
given in Sec.~\ref{sec:fit:time:swimming},
the PDF for secondary charm
may be written as
\begin{equation}
\begin{split}
  f_2(t_0, \Delta t, t_D, \lnipchisq | \mathrm{sec}) = &
    f_2(\lnipchisq | t_D ; \mathrm{sec}) \,
    f_2(t_D | t_0, \Delta t ; \mathrm{sec}) \\
    & \times f_{\mathrm{TP,0}} (t_0 | D) \,
    f_{\mathrm{TP,\Delta}} (\Delta t | D)
    .
    \label{eq:fit:time:lifetime:secondaryFullPDF}
\end{split}
\end{equation}

The combinatorial background is described in a different way.
To begin, a nonparametric distribution is fitted to the
data in the mass sidebands. However, this model,
a two-dimensional Gaussian kernel function,
cannot be used directly in the fit: the PDF used must depend
explicitly on the turning point variables~\cite{Punzi:2004wh}.
Therefore, an unfolding procedure is applied to obtain the
underlying decay time distribution before acceptance effects.
The acceptance is then incorporated in the same way as for the
other components.
The PDF for combinatorial background
may be written as
\begin{equation}
\begin{split}
  f_2(t_0, \Delta t, t_D, \lnipchisq | \mathrm{cmb}) = &
    f_2(\lnipchisq | t_D ; \mathrm{cmb}) \,
    f_2(t_D | t_0, \Delta t ; \mathrm{cmb}) \\
    & \times f_{\mathrm{TP,0}} (t_0 | \mathrm{cmb}) \,
    f_{\mathrm{TP,\Delta}} (\Delta t | \mathrm{cmb})
    ,
\end{split}
\end{equation}
where
$f_{\mathrm{TP,0}} (t_0 | \mathrm{cmb})$ and
$f_{\mathrm{TP,\Delta}} (\Delta t | \mathrm{cmb})$
are obtained as described in Sec.~\ref{sec:fit:time:swimming},
and $f_2(\lnipchisq | t_D ; \mathrm{cmb})$
and $f_2(t_D | t_0, \Delta t ; \mathrm{cmb})$ are
derived from the distributions in the mass sidebands
as described above.

Combining the above, the total PDF used in the fit is
\begin{equation}
  f_2(t_0, \Delta t, t_D, \lnipchisq) =
  \sum_j f_2(t_0, \Delta t, t_D, \lnipchisq | j) P_2(j)
  ,
\end{equation}
where the index $j$ runs over the
prompt, secondary, and combinatoric components
and $\sum_j P_2(j)=1$.
The value of $P_2(\mathrm{cmb})$ is fixed based on
the results of the preceeding fit to $m_D$.
The sum
$[ P_2(\mathrm{prm})+P_2(\mathrm{sec}) ]$
is likewise fixed, but with the
secondary fraction of the signal free.

Pseudoexperiments are used to validate the fit procedure.
In each pseudoexperiment, events from each category
(prompt \Dz mesons, secondary \Dz mesons, and combinatorial background)
are generated according to the expected distributions
and analysed following the same procedure as used for data,
including estimation of the per-event decay time acceptance
function with the swimming method.
In an ensemble of approximately 500 pseudoexperiments
generated assuming a true \Dz lifetime of $410\ps$,
the mean of the fitted values of $\tau_D$ is $409.92 \pm 0.06\fs$,
and the normalised residuals are described by a
Gaussian distribution with a mean of $0.016 \pm 0.049$
and a width of $1.03 \pm 0.04$.

\begin{figure}[tb]
\centering
\includegraphics[width = 0.7\linewidth]{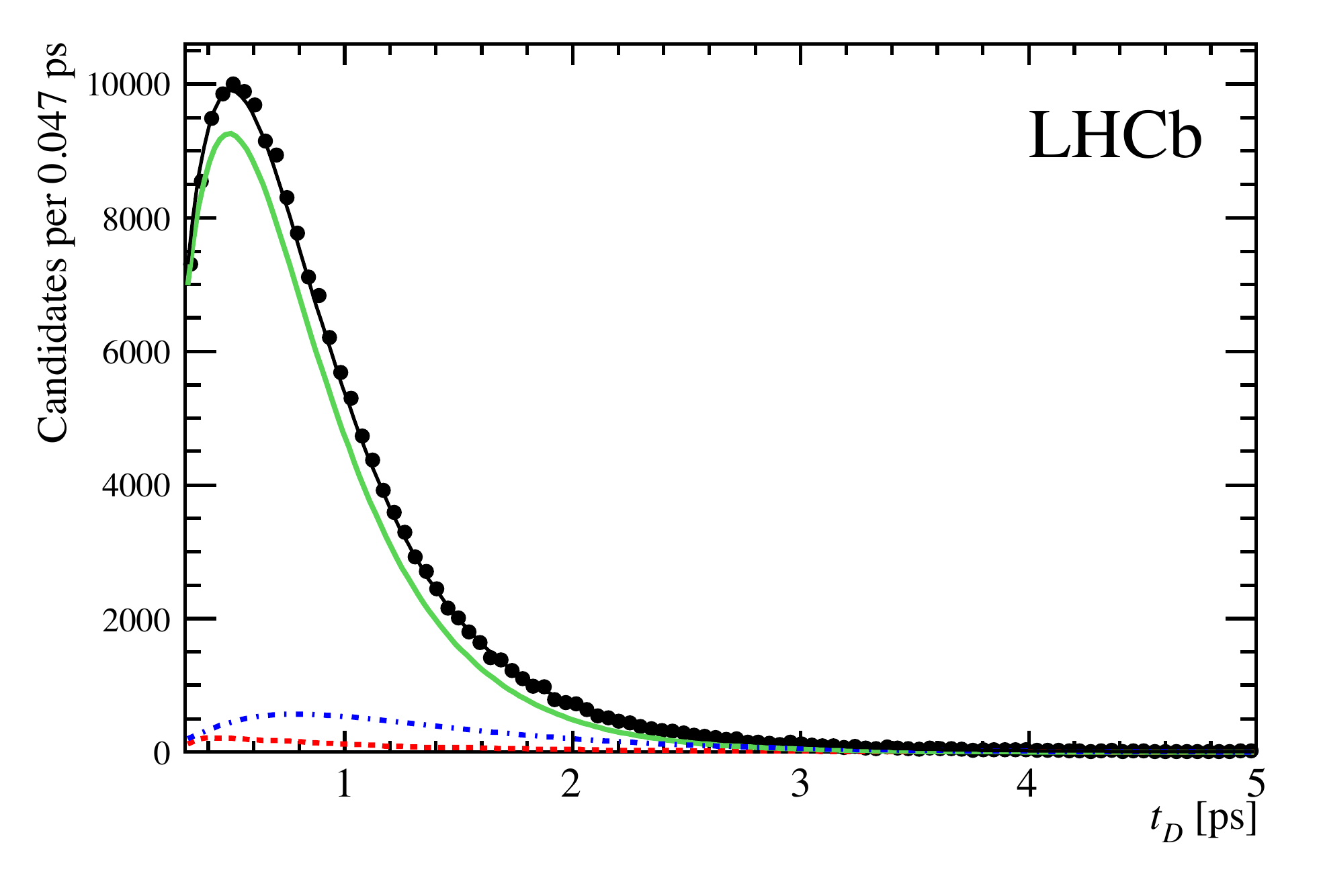} 
\includegraphics[width = 0.7\linewidth]{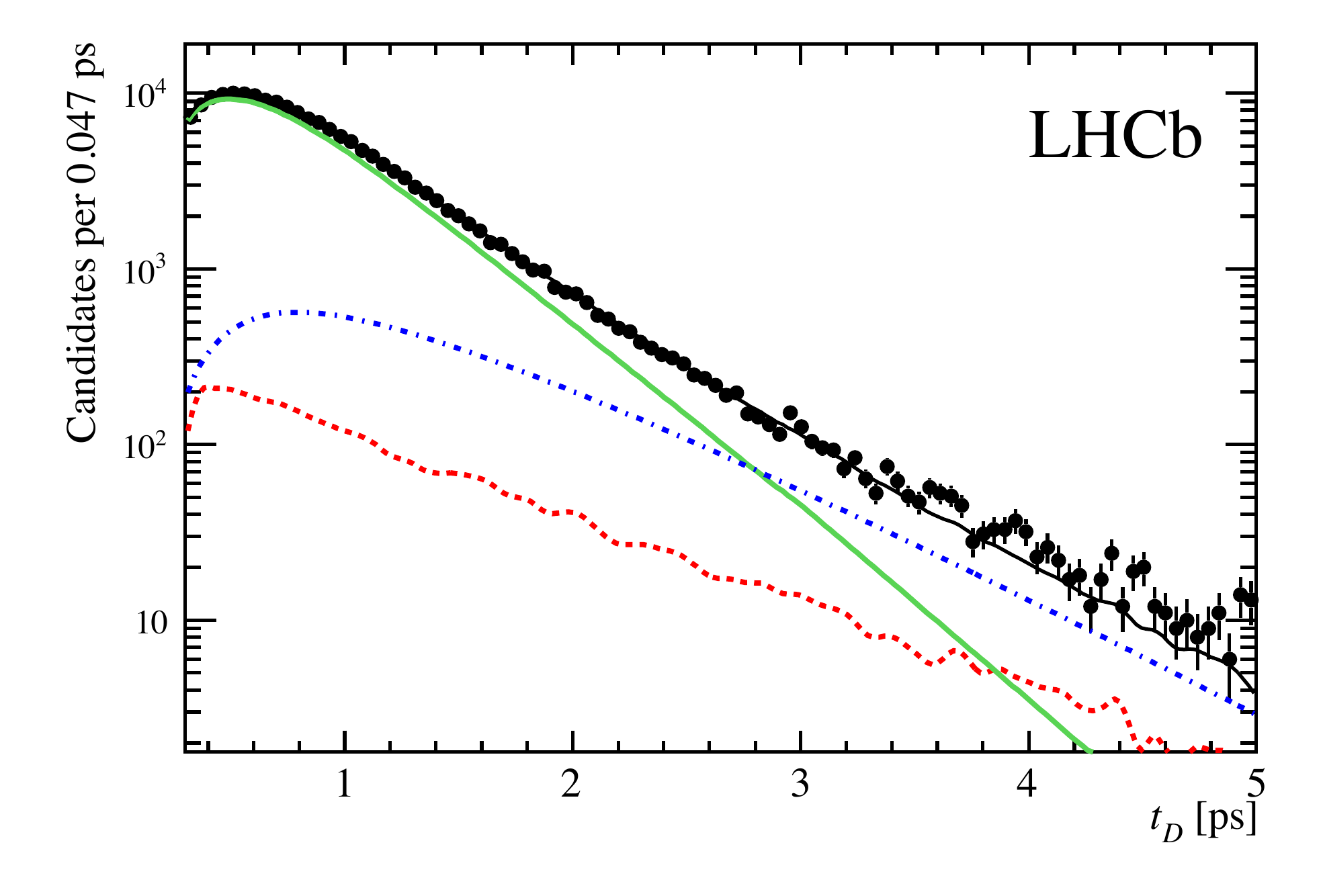} 
\caption{
  Decay time projection from the fit for separation of prompt and secondary candidates.
  The curves show the results of the fit described in
  Sec.~\ref{sec:fit:time:lifetime}:
  the total (solid black),
  the prompt component (solid green),
  the secondary component (dot-dashed blue),
  and the combinatorial component (dashed red).
  Both plots show the same data sample with linear (top) and
  logarithmic (bottom) vertical scales.
}
\label{fig:fit:time:lifetime:fitToDataDecayTime}
\end{figure}

Applying the fit to the data,
the measured lifetime is $\tau_D = 410.9 \pm 1.1 \fs$,
where the uncertainty is purely statistical. This is
consistent with the world average value of
$410.1 \pm 1.5 \fs$~\cite{PDG2014}.
The agreement between the fit and data is shown in
Fig.~\ref{fig:fit:time:lifetime:fitToDataDecayTime}.
An excess is seen at very long decay times, likely
due to imperfect modelling of the secondary component,
but there is no effect on the measurement of the lifetime of the prompt component.

\subsection{Fits to $m_D$ and $\Delta m$}
\label{sec:fit:massAndDeltaMass}

The third step consists of separate fits to the $(m_D, \Delta m)$ distributions of the
phase space bins. The fits include three components:
  $\Dstarp$ signal~(sig),
  background from genuine \Dz that are combined with an unrelated soft pion~(Dbg),
  and combinatorial background~(cmb).
In each case, the PDF is assumed to factorise into $m_D$-dependent and $\Delta m$-dependent terms.
The three components may be written as
\begin{eqnarray}
  f_3(m_D, \Delta m | \mbox{sig})  &=& f_3(m_D | \mathrm{peak})   \, f_3(\Delta m | \mathrm{peak}) , \\
  f_3(m_D, \Delta m | \mbox{Dbg}) &=& f_3(m_D | \mathrm{peak})   \, f_3(\Delta m | \mathrm{smooth}) , \\
  f_3(m_D, \Delta m | \mbox{cmb})    &=& f_3(m_D | \mathrm{smooth}) \, f_3(\Delta m | \mathrm{smooth}) ,
\end{eqnarray}
where the peaking components are defined as
\begin{eqnarray}
  f_3(m_D | \mathrm{peak}) &=& 
    \eta_1 \, G( m_D ; \mu_D, \sigma_1) + 
    \eta_2 \, G( m_D ; \mu_D, \sigma_2)
    \nonumber \\ &&
    + \, (1-\eta_1-\eta_2) \, C( m_D ; \mu_D, \sigma_3, \alpha, n) , \\
  f_3(\Delta m | \mathrm{peak}) &=&
    \eta_3 \, G(\Delta m ; \mu_{\Delta m}, \sigma_4) +
    \eta_4 \, G(\Delta m ; \mu_{\Delta m}, \sigma_5)
    \nonumber \\ &&
    + \, (1-\eta_3-\eta_4) \, B( \Delta m ; \mu_{\Delta m}, \sigma_L, \sigma_R)
  .
\end{eqnarray}
For the nonpeaking components,
  $f_3(m_D | \mathrm{smooth})$ is an exponential function and
  $f_3(\Delta m | \mathrm{smooth})$ is a second-order polynomial.
The total PDF may then be written as
\begin{equation}
  f_3(m_D, \Delta m) = \sum_j f_3(m_D, \Delta m | j) P_3(j)
  ,
\end{equation}
where the index $j$ runs over the
signal, \Dz background, and combinatoric components,
and $\sum_j P_3(j)=1$.

\begin{figure}
\centering
\includegraphics[width = 0.49\linewidth]{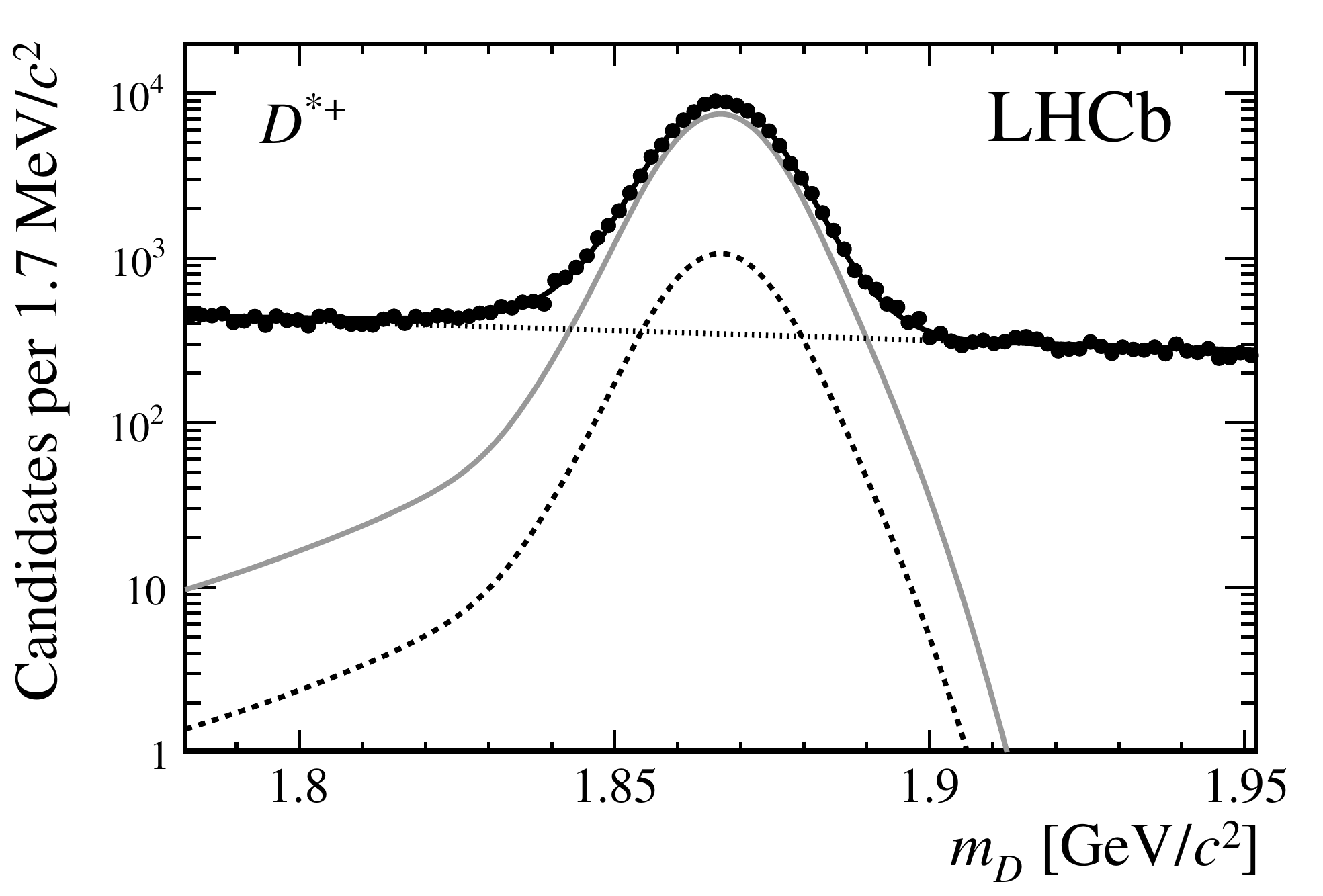}
\includegraphics[width = 0.49\linewidth]{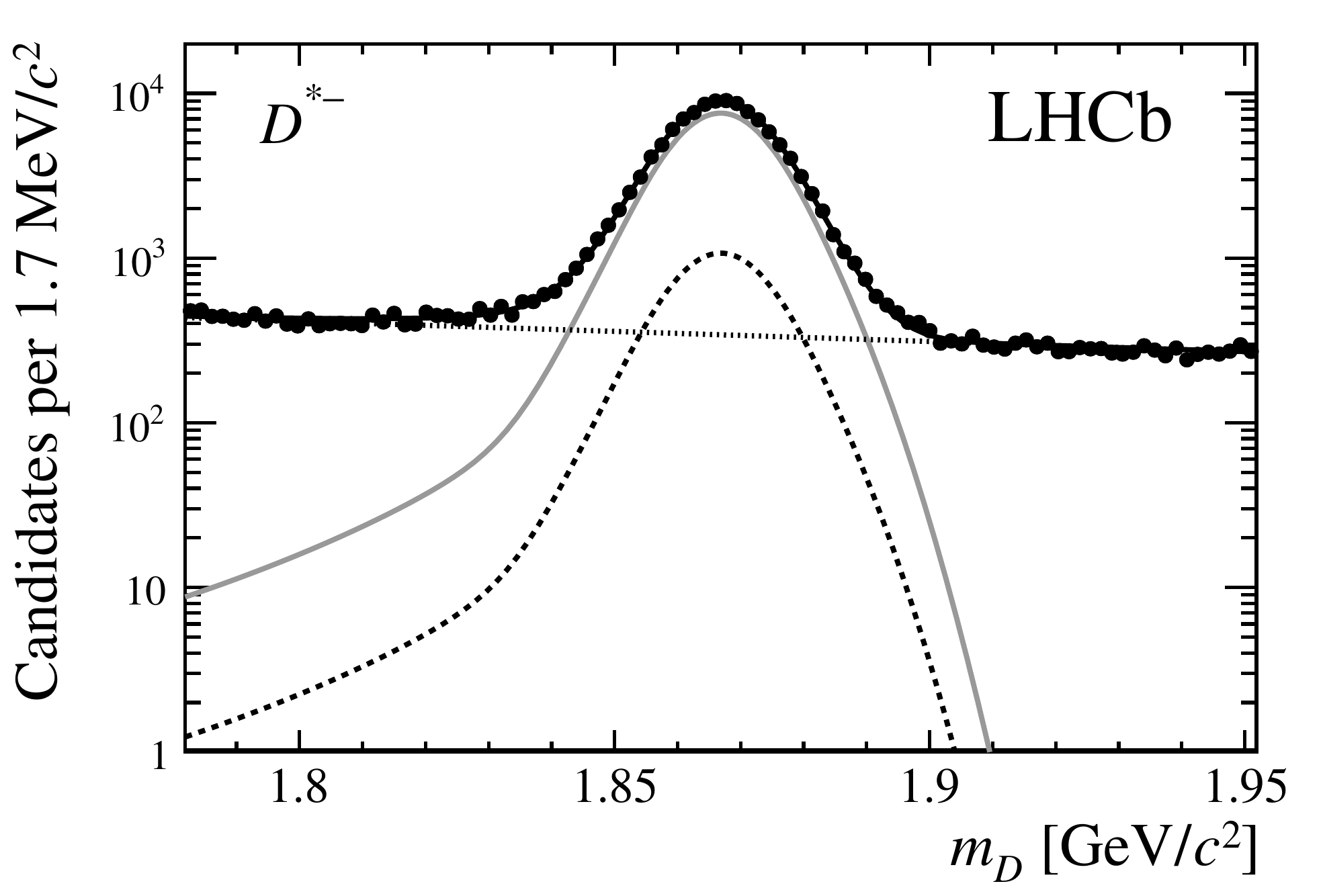}
\includegraphics[width = 0.49\linewidth]{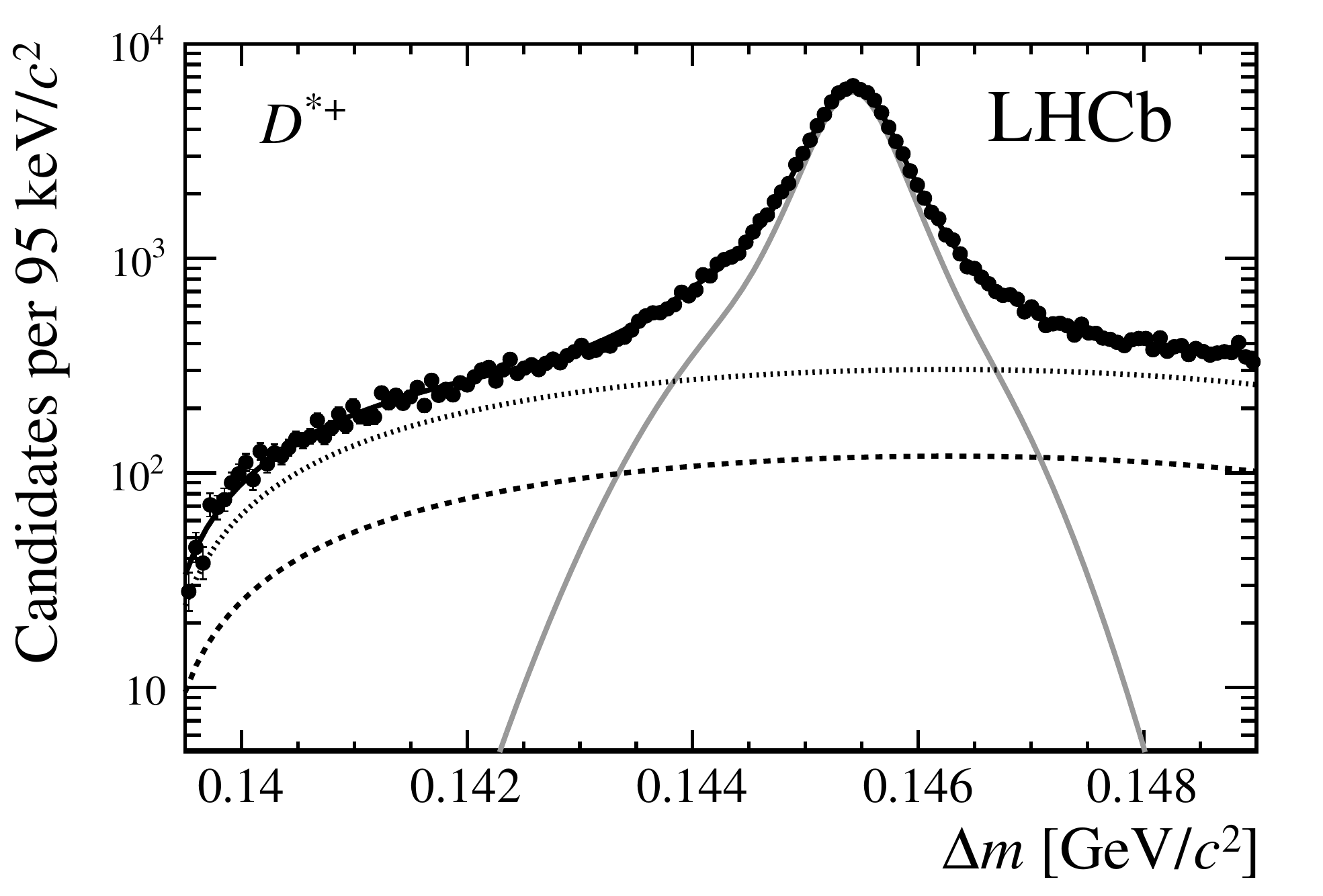}
\includegraphics[width = 0.49\linewidth]{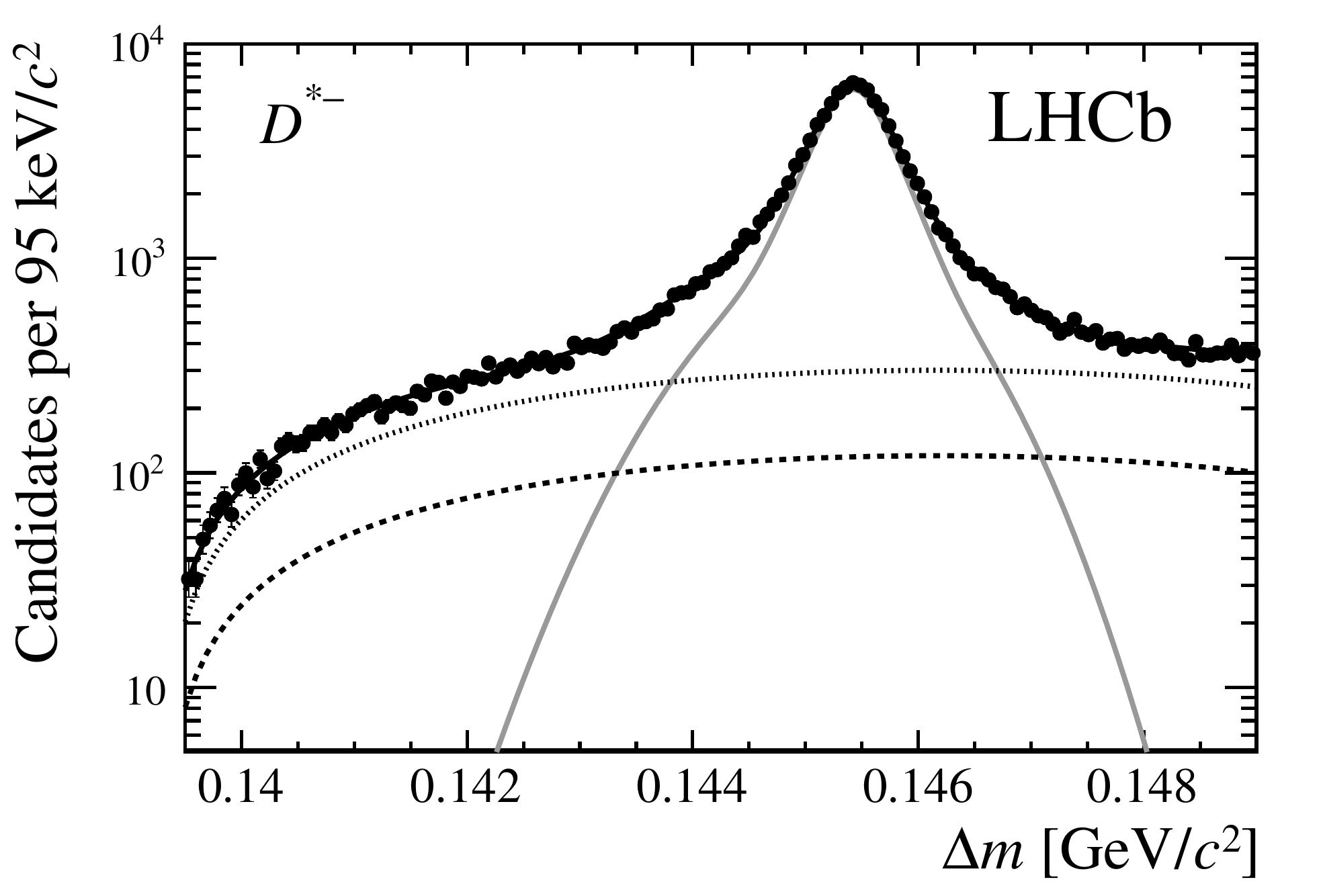}
\caption{
  Fitted $(m_D, \Delta m)$ distributions.
  The upper row shows the $m_D$ projection and the lower row $\Delta m$.
  The left column shows \Dstarp candidates and the right column \Dstarm.
  The signal and background components are shown separately
  (signal as solid grey,
   \Dz background dashed,
   combinatoric background dotted,
   and the sum as solid black).
}
\label{fig:fit:mass:fit3}
\end{figure}

To avoid an excessive number of free parameters when splitting the data into
many independent subsamples, the third fit is done in two stages.
Initially, fits to $f_3(m_D, \Delta m)$ are done without dividing the
data by phase space bin such that there are only two subsamples, \Dstarp and \Dstarm.
The results of these fits are shown in Fig.~\ref{fig:fit:mass:fit3}
and correspond to yields of approximately
85k each of \Dstarp and \Dstarm within the narrow signal window.
The parameters for
  $f_3(m_D | \mathrm{peak})$,
  $f_3(\Delta m | \mathrm{peak})$, and
  $f_3(\Delta m | \mathrm{smooth})$ are then fixed.
Individual fits to each of the 32 subsamples are then carried out, with only
the parameters of the combinatorial background shape,
  $f_3(m_D | \mathrm{smooth})$,
and the yield fractions
  $P_3(j)$
free.

\subsection{Mixing parameters}
\label{sec:fit:time:mixing}

The fourth fit uses the $(t_D, \lnipchisq)$ distributions in each
of the phase space bins for \Dz and \Dzb to determine the mixing
parameters $x$ and $y$. For a particular phase space bin $i$
and $\Dstarpm$ charge $q$, the total PDF is
\begin{equation}
  f_4(t_0, \Delta t, t_D, \lnipchisq, i, q) =
  \sum_j f_4(t_0, \Delta t, t_D, \lnipchisq, i, q | j) P_4(i,q,j)
  ,
  \label{eq:fit:time:mixing:totalPDF}
\end{equation}
where $\sum_j P_4(i,q,j)=1$
and the index $j$ runs over the
components: 
  prompt \Dstarpm (p-sig),
  prompt \Dz background (p-Dbg),
  secondary \Dstarpm (s-sig),
  secondary \Dz background (s-Dbg),
  and combinatorial background (cmb).

The prompt \Dstarpm component comprises prompt \Dz or \Dzb mesons whose
initial flavour is correctly identified. Its underlying
decay time distribution is given by
${\cal {P}}_{\Dz}(i; t_D)$ in Eq.~\ref{eq:formalism:timeDependenceForDz} for \Dstarp
and by
${\cal {P}}_{\Dzb}(i; t_D)$ in Eq.~\ref{eq:formalism:timeDependenceForDzb} for \Dstarm,
denoted ${\cal {P}}_q$.
Taking the time-dependent acceptance into account in the same way as was done for the
second fit in Eq.~\ref{eq:fit:time:lifetime:promptTimeDistributionRaw},
the per-candidate decay time PDF is
\begin{equation}
  f_4(t_D | t_0, \Delta t, i, q ; \mbox{p-sig}) = n \, a(t_D ; t_0, \Delta t) \, {\cal {P}}_q (i; t_D)
  ,
  \label{eq:fit:time:mixing:promptTimeDistributionRaw}
\end{equation}
where $n$ is a normalisation constant.
The \lnipchisq distribution for prompt \Dstarp signal at a given decay time
is fixed to that obtained in the second fit
(see Eq.~\ref{eq:fit:time:lifetime:promptlnipchisq}),
as is that for prompt \Dz background,
\begin{equation}
  f_4(\lnipchisq | t_D ; \mbox{p-sig}) = 
  f_4(\lnipchisq | t_D ; \mbox{p-Dbg}) = 
  f_2(\lnipchisq | t_D ; \mbox{prm})
  .
\end{equation}
The non-parametric turning point distributions,
$f_{\mathrm{TP,0}} (t_0 | i; D)$ and
$f_{\mathrm{TP,\Delta}} (\Delta t | i; D)$,
are obtained in the same way as was done for
the second fit, except that each phase space bin is now considered separately;
here the label $D$ denotes that the distributions are used for all components
that contain a real \Dz or \Dzb (p-sig, p-Dbg, s-sig, s-Dbg).
The prompt \Dstarp PDF is
\begin{equation}
  \begin{split}
    f_4(t_0, \Delta t, t_D, \lnipchisq, i, q | \mbox{p-sig}) =
      & f_4(\lnipchisq | t_D ; \mbox{p-sig}) \,
        f_4(t_D | t_0, \Delta t, i, q ; \mbox{p-sig}) \\
      & \times 
        f_{\mathrm{TP,0}} (t_0 |i;  D) \,
        f_{\mathrm{TP,\Delta}} (\Delta t | i; D)
      .
  \end{split}
\end{equation}

The prompt \Dz background component consists of correctly reconstructed prompt \Dz
(or \Dzb) mesons, each of which is paired with an unrelated soft pion such that the
assigned initial flavour is random. Ignoring the assigned flavour,
the underlying decay time distribution for phase space bin $i$, $u(t_D ; i)$,
is a linear combination of 
${\cal {P}}_{\Dz}(i; t_D)$ and ${\cal {P}}_{\Dzb}(i; t_D)$. The coefficients
depend on the relative populations of bin $i$ for \Dz
and bin $-i$ for \Dzb, $T_i$ and $T_{-i}$ defined in Eq.~\ref{eq:formalism:defineTi},
since the \Dzb Dalitz plot is the mirror reflection of
that of \Dz neglecting \CP violation.
The underlying decay time distribution is thus
\begin{equation}
  u(t_D ; i) =
  \frac{
    p_{\Dz} T_i {\cal {P}}_{\Dz}(i; t_D) + (1 - p_{\Dz}) T_{-i} {\cal {P}}_{\Dzb}(i; t_D)
  }{
    p_{\Dz} T_i + (1 - p_{\Dz}) T_{-i}
  }
  ,
\end{equation}
where $p_{\Dz}$ is the fraction of the prompt \Dz background due to \Dz mesons
and $(1-p_{\Dz})$ is the fraction due to \Dzb mesons.
Since production and detection charge asymmetries for pions in the relevant
kinematic region are small~\cite{LHCb-PAPER-2012-009}, $p_{\Dz}$ is assumed to be $0.5$.
The per-candidate decay time PDF is then
\begin{equation}
  f_4(t_D | t_0, \Delta t, i, q ; \mbox{p-Dbg}) = n \, a(t_D ; t_0, \Delta t) \, u(t_D ; i)
  ,
\end{equation}
where $n$ is again a calculable normalisation factor.
The turning point distributions 
$f_{\mathrm{TP,0}} (t_0 | \mathrm{prompt})$ and
$f_{\mathrm{TP,\Delta}} (\Delta t | \mathrm{prompt})$
are fixed to be the same as those obtained in the second fit.
The prompt \Dz background PDF is
\begin{equation}
  \begin{split}
    f_4(t_0, \Delta t, t_D, \lnipchisq, i, q | \mbox{p-Dbg}) =&
      f_4(\lnipchisq | t_D ; \mbox{p-Dbg}) 
      \, f_4(t_D | t_0, \Delta t, i, q ; \mbox{p-Dbg})
      \\ & \times 
      \, f_{\mathrm{TP,0}} (t_0 | i; D)
      \, f_{\mathrm{TP,\Delta}} (\Delta t | i; D)
      .
  \end{split}
\end{equation}

For the
secondary \Dstarpm
and
secondary \Dz background
components, the effect of mixing is neglected so that
the underlying time distribution does not depend on
the identified flavour or on the phase space bin.
The same functional form is used as for the second fit,
and the parameters are fixed to those obtained in the
second fit. Thus, the PDF is the same as that given
in Eq.~\ref{eq:fit:time:lifetime:secondaryFullPDF},
\begin{equation}
  f_4(t_0, \Delta t, t_D, \lnipchisq, i, q | \mbox{s-sig}) 
  = f_4(t_0, \Delta t, t_D, \lnipchisq, i, q | \mbox{s-Dbg}) 
  = f_2(t_0, \Delta t, t_D, \lnipchisq | \mathrm{sec})
  .
\end{equation}
It is assumed that the fraction of $\Dstarpm$ signal that is
from secondary production is the same in every phase space bin,
and that the same fraction also applies to the secondary \Dz
background.

For the combinatorial component,
nonparametric models are used for
the decay time and \lnipchisq distributions 
in a similar way to the second fit.
However, the distributions for each of the 32 subsamples,
split by phase space bin and by \Dstarpm charge, are modelled
independently according to the mass sidebands for that
bin and charge.

Thus, nearly all of the parameters in the total PDF
for the fourth fit (Eq.~\ref{eq:fit:time:mixing:totalPDF})
are fixed. Likewise, the fractions for each component
$P_4(i,q,j)$ are fixed based on the previous fits.
The $T_i$ values are fixed to those obtained by CLEO
(so as to reduce the number of free parameters and improve fit behaviour).
The only free parameters are
$x$, $y$, $\Gamma_D = 1/\tau_D$, and the set of $(c_i, s_i)$
values. For the latter, the information on the CLEO
measurements and their uncertainties, including 
correlations, is incorporated as a set of correlated Gaussian constraints on the likelihood. 

As in Sec.~\ref{sec:fit:time:lifetime}, pseudoexperiments are used to validate the fit procedure,
following all steps including the per-event decay time acceptance determination.
An ensemble of 1000 experiments is generated
with \cfit~\cite{cfit}
taking
$\Gamma_D = 2.44 \invps$, $x = -1\tmt$, and $y = +1\tmt$.
The mean fitted values of $x$ and $y$ are found to differ
from the input values by $(-0.016 \pm 0.014)\tmt$ and
$(+0.013 \pm 0.016)\tmt$, respectively.
The mean fitted value of $\Gamma_D$ differs from the input
value by $(+0.0012 \pm 0.0002)\invps$; although this
indicates a measureable bias, it is
only approximately
one sixth the size of the statistical uncertainty on $\Gamma_D$.
Since $\Gamma_D$ is measured here only as a cross-check,
this is ignored.
Validation tests are also performed with a sample of
pseudoevents generated with \pythia and \evtgen, corresponding
to approximately double the yield in data,
and with a sample of events in which the full detector response
was simulated with \geant, corresponding to approximately
a quarter of the yield in data. The output is
consistent with the input values of the mixing parameters supplied
to the generators.

The results of the fit to data are
\begin{eqnarray*}
  x &=& (-0.86 \pm 0.53) \tmt , \\
  y &=& (+0.03 \pm 0.46) \tmt , \\
  \Gamma_D &=& 2.435 \pm 0.006 \invps
  .
\end{eqnarray*}
The correlation coefficient between $x$ and $y$ is $+0.37$.
The uncertainties quoted above are the statistical uncertainties estimated
by the likelihood fit. They do not include any systematic effects,
but they do implicitly include the propagated uncertainties on the
CLEO $(c_i, s_i)$ parameters.
These are estimated with pseudoexperiments to be in the range $(0.05$--$0.15)\tmt$.
As a check, the fit to data is repeated with the $(c_i, s_i)$
values fixed to those obtained by CLEO, giving
$x = (-0.73 \pm 0.48)\tmt$ and $y = (+0.05 \pm 0.45)\tmt$, with $\Gamma_D$ unchanged.
The shifts in $x$ and $y$ are consistent with the uncertainties associated with
the CLEO parameters.

\section{Systematic uncertainties}
\label{sec:sys}

Further
cross-checks are performed and systematic effects considered,
as summarised in Table~\ref{tab:sys}.
Several sources of systematic uncertainty
are due to assumptions made for the baseline fit
procedure. These uncertainties are estimated
with ensembles of pseudoexperiments in which events
are generated so as to mimic the effect being studied.
For these tests, the systematic uncertainties on $x$ and $y$
are typically estimated as the sum in quadrature of the
shift in the central value and the uncertainty on the shift.
The fit procedure was also validated with a sample of
events in which the detector response was simulated using \geant
as outlined in Sec.~\ref{sec:Detector}; the values of $x$
and $y$ obtained were consistent with the input parameters.

  Biases on $x$ and $y$ due to the fit procedure itself are assessed
    through the use of pseudoexperiments.
  The resolutions on the decay time, on the turning points, and on
    $m_{12}^2$ and $m_{13}^2$ are evaluated by generating pseudoexperiments
    with resolution smearing and then fitting them
    with the baseline procedure in which the resolution is neglected.
    Estimates of the resolutions are taken from data or from the full
    simulation based on \geant.
  The assumption that the turning point distributions
    of prompt and secondary signal are equivalent is tested with
    pseudoexperiments in which these distributions are drawn from
    prompt-enriched ($\lnipchisq < 1$) and secondary-enriched
    ($\lnipchisq > 3$) samples, respectively.
  The impact of neglecting variation in efficiency as a function of
    position in the Dalitz plot is assessed by generating
    pseudoexperiments with a nonuniform efficiency model,
    determined with full simulation, and fitting them with the
    baseline procedure.
    The efficiency is described by a polynomial
    function and the following variations are tested:
      the order of the polynomial,
      whether or not it is required to be symmetric about the leading diagonal in the Dalitz plot,
      and the use of a different event selection.
    The variation among models in the values of $x$ and $y$
    is smaller than the systematic uncertainties quoted, which are
    based on the variation with respect to the baseline fit;
    in particular, the variation in $x$ among the models is approximately $0.01\tmt$.
  The uncertainty associated with the model of the tracking
    efficiency correction $\varepsilon(t_D)$,
    discussed in Sec.~\ref{sec:fit:time:swimming},
    is assessed by allowing higher-order terms in the model.
  Due to the absence of a \KS mass constraint, a small fraction of
    events fall outside the expected Dalitz plot boundary in the
    baseline procedure and an algorithm is used to assign them to
    a nearby bin; the effect of this is tested by instead rejecting
    all such events.
  To test the modelling of the combinatorial background, 
    the procedure is repeated using just the data in 
    one of the two sidebands, with the \Dstarp and \Dstarm
    samples separated (as in the baseline fit) or combined.

In addition, the uncertainties associated with a number of parameters
that are fixed in the baseline fit are included, generally by
rerunning the baseline fit repeatedly with the parameters fixed
to different values obtained by smearing the nominal values
randomly according to their estimated uncertainties.
This procedure is used
  for the $T_i$ values from CLEO,
  for the yield fractions estimated from the third fit to the
    $(m_D, \Delta m)$ distribution, and
  for the decay time and \lnipchisq parameters fixed
    based on the second fit.
The effects of varying
  the \Dz-\Dzb composition of the prompt \Dz background (via the fraction $p_{\Dz}$)
  and of using separate models of the prompt and secondary \lnipchisq
  distributions for each phase space bin
are also tested.

The sum in quadrature
of the systematic uncertainties
is $0.17\tmt$ for $x$ and $0.13\tmt$ for $y$.

\begin{table}
\caption{
  Systematic uncertainties on $x$ and $y$.
  The statistical uncertainties, which include
  the uncertainties associated with the CLEO parameters $(c_i, s_i)$,
  are shown for comparison.
}
\begin{center}
\begin{tabular}{lcc}
Source                                                  & $x \, (\tmt)$ & $y \, (\tmt)$ \\ \hline
Fit bias                                                & 0.021         & 0.020         \\
Decay time resolution                                   & 0.065         & 0.039         \\
Turning point (TP) resolution                           & 0.020         & 0.022         \\
Invariant mass resolution                               & 0.073         & 0.028         \\
Prompt/secondary TP distributions                       & 0.051         & 0.023         \\
Efficiency over phase space                             & 0.057         & 0.071         \\
Tracking efficiency parameterisation                    & 0.015         & 0.025         \\
Kinematic boundary                                      & 0.012         & 0.006         \\
Combinatorial background                                & 0.061         & 0.052         \\
Treatment of secondary \D decays                        & 0.046         & 0.025         \\
Uncertainty from $T_i$                                  & 0.079         & 0.056         \\
Uncertainties from $(m_D,\Delta m)$ fits                & 0.000         & 0.000         \\
Uncertainties from lifetime fit                         & 0.020         & 0.043         \\
\Dz background                                          & 0.001         & 0.006         \\
Variation of signal components across the phase space   & 0.013         & 0.017         \\ \hline
Total systematic uncertainty                            & 0.171         & 0.134         \\
Statistical uncertainty                                 & 0.527         & 0.463         \\
\end{tabular}
\end{center}
\label{tab:sys}
\end{table}

\section{Conclusions}
\label{sec:conc}

The charm mixing parameters $x$ and $y$ have been measured
using a novel method that does not require the use of an amplitude model
but instead uses external measurements of the strong phase
made at an $e^+ e^-$ collider running at the $\psi(3770)$ resonance~\cite{Libby:2010nu}.
A sample of $\proton \proton$ collision data 
recorded by the LHCb experiment was used,
corresponding to an integrated luminosity of $1.0 \invfb$
at a centre-of-mass energy of $7\tev$.
Neglecting \CP violation, the measured values are
\begin{eqnarray*}
  x &=& (-0.86 \pm 0.53 \pm 0.17) \tmt , \\
  y &=& (+0.03 \pm 0.46 \pm 0.13) \tmt
  .
\end{eqnarray*}
The first uncertainties are combinations of the LHCb
statistical uncertainties and those due to the CLEO measurements of the
$(c_i, s_i)$ parameters, whose effect is too small to determine
precisely from the fit but is estimated to be in the range
$(0.05$--$0.15)\tmt$. The second uncertainties are systematic.
The correlation coefficient between $x$ and $y$ for the first uncertainty is $+0.37$,
and the systematic uncertainties are considered uncorrelated.
The analysis prefers a negative value of $x$, but positive
values are not excluded.
The current HFAG world averages~\cite{HFAG} are
$x = (+0.37 \pm 0.16)\tmt$
and
$y = (+0.66 \, ^{+0.07}_{-0.10})\tmt$.

This analysis constitutes a proof of principle that
the mixing parameters can be measured
  in $\Dz \to \KS \pip \pim$ decays at LHCb
  without the need for an amplitude model.
The statistical uncertainty will be be reduced
substantially by the addition of the 2012 data sample
due to improvements in the software trigger,
which now accepts
$\Dz \to \KS \pip \pim$ decays in which the \KS vertex lies outside the
vertex detector, as occurs in the majority of cases.
A further improvement may be obtained
if charm mesons produced in semileptonic $b$-hadron decays
are incorporated.
The method does not require a detailed model of the
  efficiency as a function of position in the phase space,
  and the decay time acceptance is determined from data.
Thus, the method does not rely on the extensive use of Monte Carlo
  simulation. This is crucial for future analyses,
  especially in the context of the planned LHCb upgrade
  where $\mathcal{O}(10^8)$ signal events are expected~\cite{LHCb-PAPER-2012-031}.
  To take full advantage of such a data set, more precise
  strong phase measurements from a charm factory
  running on the $\psi(3770)$ resonance will be needed.

\section*{Acknowledgements}

\noindent We express our gratitude to our colleagues in the CERN
accelerator departments for the excellent performance of the LHC. We
thank the technical and administrative staff at the LHCb
institutes. We acknowledge support from CERN and from the national
agencies: CAPES, CNPq, FAPERJ and FINEP (Brazil); NSFC (China);
CNRS/IN2P3 (France); BMBF, DFG and MPG (Germany); INFN (Italy); 
FOM and NWO (The Netherlands); MNiSW and NCN (Poland); MEN/IFA (Romania); 
MinES and FANO (Russia); MinECo (Spain); SNSF and SER (Switzerland); 
NASU (Ukraine); STFC (United Kingdom); NSF (USA).
We acknowledge the computing resources that are provided by CERN, IN2P3 (France), KIT and DESY (Germany), INFN (Italy), SURF (The Netherlands), PIC (Spain), GridPP (United Kingdom), RRCKI (Russia), CSCS (Switzerland), IFIN-HH (Romania), CBPF (Brazil), PL-GRID (Poland) and OSC (USA). We are indebted to the communities behind the multiple open 
source software packages on which we depend. We are also thankful for the 
computing resources and the access to software R\&D tools provided by Yandex LLC (Russia).
Individual groups or members have received support from AvH Foundation (Germany),
EPLANET, Marie Sk\l{}odowska-Curie Actions and ERC (European Union), 
Conseil G\'{e}n\'{e}ral de Haute-Savoie, Labex ENIGMASS and OCEVU, 
R\'{e}gion Auvergne (France), RFBR (Russia), XuntaGal and GENCAT (Spain), The Royal Society 
and Royal Commission for the Exhibition of 1851 (United Kingdom).



\addcontentsline{toc}{section}{References}
\setboolean{inbibliography}{true}
\bibliographystyle{LHCb}
\bibliography{main,LHCb-PAPER,LHCb-CONF,LHCb-DP,LHCb-TDR,kshh-refs}

\ifx\mcitethebibliography\mciteundefinedmacro
\PackageError{LHCb.bst}{mciteplus.sty has not been loaded}
{This bibstyle requires the use of the mciteplus package.}\fi
\providecommand{\href}[2]{#2}
\begin{mcitethebibliography}{10}
\mciteSetBstSublistMode{n}
\mciteSetBstMaxWidthForm{subitem}{\alph{mcitesubitemcount})}
\mciteSetBstSublistLabelBeginEnd{\mcitemaxwidthsubitemform\space}
{\relax}{\relax}

\bibitem{LHCb-PAPER-2012-038}
LHCb collaboration, R.~Aaij {\em et~al.},
  \ifthenelse{\boolean{articletitles}}{\emph{{Observation of
  $D^0$--$\overline{D}^0$ oscillations}},
  }{}\href{http://dx.doi.org/10.1103/PhysRevLett.110.101802}{Phys.\ Rev.\
  Lett.\  \textbf{110} (2013) 101802},
  \href{http://arxiv.org/abs/1211.1230}{{\normalfont\ttfamily
  arXiv:1211.1230}}\relax
\mciteBstWouldAddEndPuncttrue
\mciteSetBstMidEndSepPunct{\mcitedefaultmidpunct}
{\mcitedefaultendpunct}{\mcitedefaultseppunct}\relax
\EndOfBibitem
\bibitem{Aaltonen:2013pja}
CDF collaboration, T.~A. Aaltonen {\em et~al.},
  \ifthenelse{\boolean{articletitles}}{\emph{{Observation of \Dz-\Dzb mixing
  using the CDF II detector}},
  }{}\href{http://dx.doi.org/10.1103/PhysRevLett.111.231802}{Phys.\ Rev.\
  Lett.\  \textbf{111} (2013) 231802},
  \href{http://arxiv.org/abs/1309.4078}{{\normalfont\ttfamily
  arXiv:1309.4078}}\relax
\mciteBstWouldAddEndPuncttrue
\mciteSetBstMidEndSepPunct{\mcitedefaultmidpunct}
{\mcitedefaultendpunct}{\mcitedefaultseppunct}\relax
\EndOfBibitem
\bibitem{LHCb-PAPER-2013-053}
LHCb collaboration, R.~Aaij {\em et~al.},
  \ifthenelse{\boolean{articletitles}}{\emph{{Measurement of
  $D^0$--$\overline{D}^0$ mixing parameters and search for $C\!P$ violation
  using $D^0\to K^+\pi^-$ decays}},
  }{}\href{http://dx.doi.org/10.1103/PhysRevLett.111.251801}{Phys.\ Rev.\
  Lett.\  \textbf{111} (2013) 251801},
  \href{http://arxiv.org/abs/1309.6534}{{\normalfont\ttfamily
  arXiv:1309.6534}}\relax
\mciteBstWouldAddEndPuncttrue
\mciteSetBstMidEndSepPunct{\mcitedefaultmidpunct}
{\mcitedefaultendpunct}{\mcitedefaultseppunct}\relax
\EndOfBibitem
\bibitem{Ko:2014qvu}
Belle collaboration, B.~R. Ko {\em et~al.},
  \ifthenelse{\boolean{articletitles}}{\emph{{Observation of \Dz-\Dzb mixing in
  $e^+e^-$ collisions}},
  }{}\href{http://dx.doi.org/10.1103/PhysRevLett.112.111801}{Phys.\ Rev.\
  Lett.\  \textbf{112} (2014) 111801},
  \href{http://arxiv.org/abs/1401.3402}{{\normalfont\ttfamily
  arXiv:1401.3402}}\relax
\mciteBstWouldAddEndPuncttrue
\mciteSetBstMidEndSepPunct{\mcitedefaultmidpunct}
{\mcitedefaultendpunct}{\mcitedefaultseppunct}\relax
\EndOfBibitem
\bibitem{HFAG}
Heavy Flavor Averaging Group, Y.~Amhis {\em et~al.},
  \ifthenelse{\boolean{articletitles}}{\emph{{Averages of $b$-hadron,
  $c$-hadron, and $\tau$-lepton properties as of summer 2014}},
  }{}\href{http://arxiv.org/abs/1412.7515}{{\normalfont\ttfamily
  arXiv:1412.7515}}, {updated results and plots available at
  \href{http://www.slac.stanford.edu/xorg/hfag/}{{\tt
  http://www.slac.stanford.edu/xorg/hfag/}}}\relax
\mciteBstWouldAddEndPuncttrue
\mciteSetBstMidEndSepPunct{\mcitedefaultmidpunct}
{\mcitedefaultendpunct}{\mcitedefaultseppunct}\relax
\EndOfBibitem
\bibitem{Asner:2005sz}
CLEO collaboration, D.~M. Asner {\em et~al.},
  \ifthenelse{\boolean{articletitles}}{\emph{{Search for \Dz-\Dzb mixing in the
  Dalitz plot analysis of $\Dz \to \KS \pip \pim$}},
  }{}\href{http://dx.doi.org/10.1103/PhysRevD.72.012001}{Phys.\ Rev.\
  \textbf{D72} (2005) 012001},
  \href{http://arxiv.org/abs/hep-ex/0503045}{{\normalfont\ttfamily
  arXiv:hep-ex/0503045}}\relax
\mciteBstWouldAddEndPuncttrue
\mciteSetBstMidEndSepPunct{\mcitedefaultmidpunct}
{\mcitedefaultendpunct}{\mcitedefaultseppunct}\relax
\EndOfBibitem
\bibitem{delAmoSanchez:2010xz}
BaBar collaboration, P.~del Amo~Sanchez {\em et~al.},
  \ifthenelse{\boolean{articletitles}}{\emph{{Measurement of \Dz-\Dzb mixing
  parameters using $\Dz \to \KS \pip \pim$ and $\Dz \to \KS \Kp \Km$ decays}},
  }{}\href{http://dx.doi.org/10.1103/PhysRevLett.105.081803}{Phys.\ Rev.\
  Lett.\  \textbf{105} (2010) 081803},
  \href{http://arxiv.org/abs/1004.5053}{{\normalfont\ttfamily
  arXiv:1004.5053}}\relax
\mciteBstWouldAddEndPuncttrue
\mciteSetBstMidEndSepPunct{\mcitedefaultmidpunct}
{\mcitedefaultendpunct}{\mcitedefaultseppunct}\relax
\EndOfBibitem
\bibitem{Peng:2014oda}
Belle collaboration, T.~Peng {\em et~al.},
  \ifthenelse{\boolean{articletitles}}{\emph{{Measurement of \Dz-\Dzb mixing
  and search for indirect CP violation using $\Dz \to \KS \pi^+ \pi^-$
  decays}}, }{}\href{http://dx.doi.org/10.1103/PhysRevD.89.091103}{Phys.\ Rev.\
   \textbf{D89} (2014) 091103},
  \href{http://arxiv.org/abs/1404.2412}{{\normalfont\ttfamily
  arXiv:1404.2412}}\relax
\mciteBstWouldAddEndPuncttrue
\mciteSetBstMidEndSepPunct{\mcitedefaultmidpunct}
{\mcitedefaultendpunct}{\mcitedefaultseppunct}\relax
\EndOfBibitem
\bibitem{Muramatsu:2002jp}
CLEO collaboration, H.~Muramatsu {\em et~al.},
  \ifthenelse{\boolean{articletitles}}{\emph{{Dalitz analysis of $\Dz \to \KS
  \pip \pim$}},
  }{}\href{http://dx.doi.org/10.1103/PhysRevLett.89.251802}{Phys.\ Rev.\ Lett.\
   \textbf{89} (2002) 251802},
  \href{http://arxiv.org/abs/hep-ex/0207067}{{\normalfont\ttfamily
  arXiv:hep-ex/0207067}}\relax
\mciteBstWouldAddEndPuncttrue
\mciteSetBstMidEndSepPunct{\mcitedefaultmidpunct}
{\mcitedefaultendpunct}{\mcitedefaultseppunct}\relax
\EndOfBibitem
\bibitem{Aaltonen:2012nd}
CDF collaboration, T.~Aaltonen {\em et~al.},
  \ifthenelse{\boolean{articletitles}}{\emph{{Measurement of CP-violation
  asymmetries in $\Dz \to \KS \pi^+ \pi^-$}},
  }{}\href{http://dx.doi.org/10.1103/PhysRevD.86.032007}{Phys.\ Rev.\
  \textbf{D86} (2012) 032007},
  \href{http://arxiv.org/abs/1207.0825}{{\normalfont\ttfamily
  arXiv:1207.0825}}\relax
\mciteBstWouldAddEndPuncttrue
\mciteSetBstMidEndSepPunct{\mcitedefaultmidpunct}
{\mcitedefaultendpunct}{\mcitedefaultseppunct}\relax
\EndOfBibitem
\bibitem{Bondar:2010qs}
A.~Bondar, A.~Poluektov, and V.~Vorobiev,
  \ifthenelse{\boolean{articletitles}}{\emph{{Charm mixing in the
  model-independent analysis of correlated \Dz \Dzb decays}},
  }{}\href{http://dx.doi.org/10.1103/PhysRevD.82.034033}{Phys.\ Rev.\
  \textbf{D82} (2010) 034033},
  \href{http://arxiv.org/abs/1004.2350}{{\normalfont\ttfamily
  arXiv:1004.2350}}\relax
\mciteBstWouldAddEndPuncttrue
\mciteSetBstMidEndSepPunct{\mcitedefaultmidpunct}
{\mcitedefaultendpunct}{\mcitedefaultseppunct}\relax
\EndOfBibitem
\bibitem{Giri:2003ty}
A.~Giri, Y.~Grossman, A.~Soffer, and J.~Zupan,
  \ifthenelse{\boolean{articletitles}}{\emph{{Determining $\gamma$ using
  $B^{\pm} \to D K^{\pm}$ with multibody $D$ decays}},
  }{}\href{http://dx.doi.org/10.1103/PhysRevD.68.054018}{Phys.\ Rev.\
  \textbf{D68} (2003) 054018},
  \href{http://arxiv.org/abs/hep-ph/0303187}{{\normalfont\ttfamily
  arXiv:hep-ph/0303187}}\relax
\mciteBstWouldAddEndPuncttrue
\mciteSetBstMidEndSepPunct{\mcitedefaultmidpunct}
{\mcitedefaultendpunct}{\mcitedefaultseppunct}\relax
\EndOfBibitem
\bibitem{Libby:2010nu}
CLEO collaboration, J.~Libby {\em et~al.},
  \ifthenelse{\boolean{articletitles}}{\emph{{Model-independent determination
  of the strong-phase difference between \Dz and $\Dzb \to K^0_{S,L} h^+ h^-$
  ($h=\pi,K$) and its impact on the measurement of the CKM angle
  $\gamma/\phi_3$}},
  }{}\href{http://dx.doi.org/10.1103/PhysRevD.82.112006}{Phys.\ Rev.\
  \textbf{D82} (2010) 112006},
  \href{http://arxiv.org/abs/1010.2817}{{\normalfont\ttfamily
  arXiv:1010.2817}}\relax
\mciteBstWouldAddEndPuncttrue
\mciteSetBstMidEndSepPunct{\mcitedefaultmidpunct}
{\mcitedefaultendpunct}{\mcitedefaultseppunct}\relax
\EndOfBibitem
\bibitem{Ablikim:2009aa}
BESIII collaboration, M.~Ablikim {\em et~al.},
  \ifthenelse{\boolean{articletitles}}{\emph{{Design and construction of the
  BESIII detector}},
  }{}\href{http://dx.doi.org/10.1016/j.nima.2009.12.050}{Nucl.\ Instrum.\
  Meth.\  \textbf{A614} (2010) 345},
  \href{http://arxiv.org/abs/0911.4960}{{\normalfont\ttfamily
  arXiv:0911.4960}}\relax
\mciteBstWouldAddEndPuncttrue
\mciteSetBstMidEndSepPunct{\mcitedefaultmidpunct}
{\mcitedefaultendpunct}{\mcitedefaultseppunct}\relax
\EndOfBibitem
\bibitem{Thomas:2012qf}
C.~Thomas and G.~Wilkinson,
  \ifthenelse{\boolean{articletitles}}{\emph{{Model-independent \Dz-\Dzb mixing
  and CP violation studies with $\Dz \to \KS \pi^+\pi^-$ and $\Dz \to \KS
  K^+K^-$}}, }{}\href{http://dx.doi.org/10.1007/JHEP10(2012)185}{JHEP
  \textbf{10} (2012) 185},
  \href{http://arxiv.org/abs/1209.0172}{{\normalfont\ttfamily
  arXiv:1209.0172}}\relax
\mciteBstWouldAddEndPuncttrue
\mciteSetBstMidEndSepPunct{\mcitedefaultmidpunct}
{\mcitedefaultendpunct}{\mcitedefaultseppunct}\relax
\EndOfBibitem
\bibitem{Alves:2008zz}
LHCb collaboration, A.~A. Alves~Jr.\ {\em et~al.},
  \ifthenelse{\boolean{articletitles}}{\emph{{The \lhcb detector at the LHC}},
  }{}\href{http://dx.doi.org/10.1088/1748-0221/3/08/S08005}{JINST \textbf{3}
  (2008) S08005}\relax
\mciteBstWouldAddEndPuncttrue
\mciteSetBstMidEndSepPunct{\mcitedefaultmidpunct}
{\mcitedefaultendpunct}{\mcitedefaultseppunct}\relax
\EndOfBibitem
\bibitem{LHCb-DP-2014-002}
LHCb collaboration, R.~Aaij {\em et~al.},
  \ifthenelse{\boolean{articletitles}}{\emph{{LHCb detector performance}},
  }{}\href{http://dx.doi.org/10.1142/S0217751X15300227}{Int.\ J.\ Mod.\ Phys.\
  \textbf{A30} (2015) 1530022},
  \href{http://arxiv.org/abs/1412.6352}{{\normalfont\ttfamily
  arXiv:1412.6352}}\relax
\mciteBstWouldAddEndPuncttrue
\mciteSetBstMidEndSepPunct{\mcitedefaultmidpunct}
{\mcitedefaultendpunct}{\mcitedefaultseppunct}\relax
\EndOfBibitem
\bibitem{LHCb-DP-2012-004}
R.~Aaij {\em et~al.}, \ifthenelse{\boolean{articletitles}}{\emph{{The \lhcb
  trigger and its performance in 2011}},
  }{}\href{http://dx.doi.org/10.1088/1748-0221/8/04/P04022}{JINST \textbf{8}
  (2013) P04022}, \href{http://arxiv.org/abs/1211.3055}{{\normalfont\ttfamily
  arXiv:1211.3055}}\relax
\mciteBstWouldAddEndPuncttrue
\mciteSetBstMidEndSepPunct{\mcitedefaultmidpunct}
{\mcitedefaultendpunct}{\mcitedefaultseppunct}\relax
\EndOfBibitem
\bibitem{PDG2014}
Particle Data Group, K.~A. Olive {\em et~al.},
  \ifthenelse{\boolean{articletitles}}{\emph{{\href{http://pdg.lbl.gov/}{Review
  of particle physics}}},
  }{}\href{http://dx.doi.org/10.1088/1674-1137/38/9/090001}{Chin.\ Phys.\
  \textbf{C38} (2014) 090001}\relax
\mciteBstWouldAddEndPuncttrue
\mciteSetBstMidEndSepPunct{\mcitedefaultmidpunct}
{\mcitedefaultendpunct}{\mcitedefaultseppunct}\relax
\EndOfBibitem
\bibitem{Sjostrand:2006za}
T.~Sj\"{o}strand, S.~Mrenna, and P.~Skands,
  \ifthenelse{\boolean{articletitles}}{\emph{{PYTHIA 6.4 physics and manual}},
  }{}\href{http://dx.doi.org/10.1088/1126-6708/2006/05/026}{JHEP \textbf{05}
  (2006) 026}, \href{http://arxiv.org/abs/hep-ph/0603175}{{\normalfont\ttfamily
  arXiv:hep-ph/0603175}}\relax
\mciteBstWouldAddEndPuncttrue
\mciteSetBstMidEndSepPunct{\mcitedefaultmidpunct}
{\mcitedefaultendpunct}{\mcitedefaultseppunct}\relax
\EndOfBibitem
\bibitem{LHCb-PROC-2010-056}
I.~Belyaev {\em et~al.}, \ifthenelse{\boolean{articletitles}}{\emph{{Handling
  of the generation of primary events in Gauss, the LHCb simulation
  framework}}, }{}\href{http://dx.doi.org/10.1088/1742-6596/331/3/032047}{{J.\
  Phys.\ Conf.\ Ser.\ } \textbf{331} (2011) 032047}\relax
\mciteBstWouldAddEndPuncttrue
\mciteSetBstMidEndSepPunct{\mcitedefaultmidpunct}
{\mcitedefaultendpunct}{\mcitedefaultseppunct}\relax
\EndOfBibitem
\bibitem{Lange:2001uf}
D.~J. Lange, \ifthenelse{\boolean{articletitles}}{\emph{{The EvtGen particle
  decay simulation package}},
  }{}\href{http://dx.doi.org/10.1016/S0168-9002(01)00089-4}{Nucl.\ Instrum.\
  Meth.\  \textbf{A462} (2001) 152}\relax
\mciteBstWouldAddEndPuncttrue
\mciteSetBstMidEndSepPunct{\mcitedefaultmidpunct}
{\mcitedefaultendpunct}{\mcitedefaultseppunct}\relax
\EndOfBibitem
\bibitem{Golonka:2005pn}
P.~Golonka and Z.~Was, \ifthenelse{\boolean{articletitles}}{\emph{{PHOTOS Monte
  Carlo: A precision tool for QED corrections in $Z$ and $W$ decays}},
  }{}\href{http://dx.doi.org/10.1140/epjc/s2005-02396-4}{Eur.\ Phys.\ J.\
  \textbf{C45} (2006) 97},
  \href{http://arxiv.org/abs/hep-ph/0506026}{{\normalfont\ttfamily
  arXiv:hep-ph/0506026}}\relax
\mciteBstWouldAddEndPuncttrue
\mciteSetBstMidEndSepPunct{\mcitedefaultmidpunct}
{\mcitedefaultendpunct}{\mcitedefaultseppunct}\relax
\EndOfBibitem
\bibitem{Allison:2006ve}
Geant4 collaboration, J.~Allison {\em et~al.},
  \ifthenelse{\boolean{articletitles}}{\emph{{Geant4 developments and
  applications}}, }{}\href{http://dx.doi.org/10.1109/TNS.2006.869826}{IEEE
  Trans.\ Nucl.\ Sci.\  \textbf{53} (2006) 270}\relax
\mciteBstWouldAddEndPuncttrue
\mciteSetBstMidEndSepPunct{\mcitedefaultmidpunct}
{\mcitedefaultendpunct}{\mcitedefaultseppunct}\relax
\EndOfBibitem
\bibitem{Agostinelli:2002hh}
Geant4 collaboration, S.~Agostinelli {\em et~al.},
  \ifthenelse{\boolean{articletitles}}{\emph{{Geant4: A simulation toolkit}},
  }{}\href{http://dx.doi.org/10.1016/S0168-9002(03)01368-8}{Nucl.\ Instrum.\
  Meth.\  \textbf{A506} (2003) 250}\relax
\mciteBstWouldAddEndPuncttrue
\mciteSetBstMidEndSepPunct{\mcitedefaultmidpunct}
{\mcitedefaultendpunct}{\mcitedefaultseppunct}\relax
\EndOfBibitem
\bibitem{LHCb-PROC-2011-006}
M.~Clemencic {\em et~al.}, \ifthenelse{\boolean{articletitles}}{\emph{{The
  \lhcb simulation application, Gauss: Design, evolution and experience}},
  }{}\href{http://dx.doi.org/10.1088/1742-6596/331/3/032023}{{J.\ Phys.\ Conf.\
  Ser.\ } \textbf{331} (2011) 032023}\relax
\mciteBstWouldAddEndPuncttrue
\mciteSetBstMidEndSepPunct{\mcitedefaultmidpunct}
{\mcitedefaultendpunct}{\mcitedefaultseppunct}\relax
\EndOfBibitem
\bibitem{Skwarnicki:1986xj}
T.~Skwarnicki, {\em {A study of the radiative cascade transitions between the
  Upsilon-prime and Upsilon resonances}}, PhD thesis, Institute of Nuclear
  Physics, Krakow, 1986,
  {\href{http://inspirehep.net/record/230779/}{DESY-F31-86-02}}\relax
\mciteBstWouldAddEndPuncttrue
\mciteSetBstMidEndSepPunct{\mcitedefaultmidpunct}
{\mcitedefaultendpunct}{\mcitedefaultseppunct}\relax
\EndOfBibitem
\bibitem{bib:swimming}
V.~V. Gligorov {\em et~al.},
  \ifthenelse{\boolean{articletitles}}{\emph{Swimming: A data driven acceptance
  correction algorithm},
  }{}\href{http://dx.doi.org/10.1088/1742-6596/396/2/022016}{J.\ Phys.\ Conf.\
  Ser.\  \textbf{396} (2012) 022016}\relax
\mciteBstWouldAddEndPuncttrue
\mciteSetBstMidEndSepPunct{\mcitedefaultmidpunct}
{\mcitedefaultendpunct}{\mcitedefaultseppunct}\relax
\EndOfBibitem
\bibitem{LHCb-PAPER-2011-032}
LHCb collaboration, R.~Aaij {\em et~al.},
  \ifthenelse{\boolean{articletitles}}{\emph{{Measurement of mixing and $C\!P$
  violation parameters in two-body charm decays}},
  }{}\href{http://dx.doi.org/10.1007/JHEP04(2012)129}{JHEP \textbf{04} (2012)
  129}, \href{http://arxiv.org/abs/1112.4698}{{\normalfont\ttfamily
  arXiv:1112.4698}}\relax
\mciteBstWouldAddEndPuncttrue
\mciteSetBstMidEndSepPunct{\mcitedefaultmidpunct}
{\mcitedefaultendpunct}{\mcitedefaultseppunct}\relax
\EndOfBibitem
\bibitem{LHCb-PAPER-2013-054}
LHCb collaboration, R.~Aaij {\em et~al.},
  \ifthenelse{\boolean{articletitles}}{\emph{{Measurements of indirect $C\!P$
  asymmetries in $D^0\to K^-K^+$ and $D^0\to\pi^-\pi^+$ decays}},
  }{}\href{http://dx.doi.org/10.1103/PhysRevLett.112.041801}{Phys.\ Rev.\
  Lett.\  \textbf{112} (2014) 041801},
  \href{http://arxiv.org/abs/1310.7201}{{\normalfont\ttfamily
  arXiv:1310.7201}}\relax
\mciteBstWouldAddEndPuncttrue
\mciteSetBstMidEndSepPunct{\mcitedefaultmidpunct}
{\mcitedefaultendpunct}{\mcitedefaultseppunct}\relax
\EndOfBibitem
\bibitem{Bailey:1985zz}
NA11 collaboration, R.~Bailey {\em et~al.},
  \ifthenelse{\boolean{articletitles}}{\emph{{Measurement of the lifetime of
  charged and neutral D mesons with high resolution silicon strip detectors}},
  }{}\href{http://dx.doi.org/10.1007/BF01413598}{Z.\ Phys.\  \textbf{C28}
  (1985) 357}\relax
\mciteBstWouldAddEndPuncttrue
\mciteSetBstMidEndSepPunct{\mcitedefaultmidpunct}
{\mcitedefaultendpunct}{\mcitedefaultseppunct}\relax
\EndOfBibitem
\bibitem{Adam:1995mb}
DELPHI collaboration, W.~Adam {\em et~al.},
  \ifthenelse{\boolean{articletitles}}{\emph{{Lifetime of charged and neutral B
  hadrons using event topology}},
  }{}\href{http://dx.doi.org/10.1007/BF01620712}{Z.\ Phys.\  \textbf{C68}
  (1995) 363}\relax
\mciteBstWouldAddEndPuncttrue
\mciteSetBstMidEndSepPunct{\mcitedefaultmidpunct}
{\mcitedefaultendpunct}{\mcitedefaultseppunct}\relax
\EndOfBibitem
\bibitem{Rademacker:2005ay}
J.~Rademacker, \ifthenelse{\boolean{articletitles}}{\emph{{Reduction of
  statistical power per event due to upper lifetime cuts in lifetime
  measurements}}, }{}\href{http://dx.doi.org/10.1016/j.nima.2006.09.090}{Nucl.\
  Instrum.\ Meth.\  \textbf{A570} (2007) 525},
  \href{http://arxiv.org/abs/hep-ex/0502042}{{\normalfont\ttfamily
  arXiv:hep-ex/0502042}}\relax
\mciteBstWouldAddEndPuncttrue
\mciteSetBstMidEndSepPunct{\mcitedefaultmidpunct}
{\mcitedefaultendpunct}{\mcitedefaultseppunct}\relax
\EndOfBibitem
\bibitem{Aaltonen:2010ta}
CDF collaboration, T.~Aaltonen {\em et~al.},
  \ifthenelse{\boolean{articletitles}}{\emph{{Measurement of the $B^-$ lifetime
  using a simulation free approach for trigger bias correction}},
  }{}\href{http://dx.doi.org/10.1103/PhysRevD.83.032008}{Phys.\ Rev.\
  \textbf{D83} (2011) 032008},
  \href{http://arxiv.org/abs/1004.4855}{{\normalfont\ttfamily
  arXiv:1004.4855}}\relax
\mciteBstWouldAddEndPuncttrue
\mciteSetBstMidEndSepPunct{\mcitedefaultmidpunct}
{\mcitedefaultendpunct}{\mcitedefaultseppunct}\relax
\EndOfBibitem
\bibitem{bib:DensityEstimation}
D.~Scott, {\em Multivariate density estimation: Theory, practice, and
  visualization}, John Wiley and Sons, Inc, 1992\relax
\mciteBstWouldAddEndPuncttrue
\mciteSetBstMidEndSepPunct{\mcitedefaultmidpunct}
{\mcitedefaultendpunct}{\mcitedefaultseppunct}\relax
\EndOfBibitem
\bibitem{Gershon:2015xra}
T.~Gershon, J.~Libby, and G.~Wilkinson,
  \ifthenelse{\boolean{articletitles}}{\emph{{Contributions to the width
  difference in the neutral $D$ system from hadronic decays}},
  }{}\href{http://dx.doi.org/10.1016/j.physletb.2015.08.063}{Phys.\ Lett.\
  \textbf{B750} (2015) 338},
  \href{http://arxiv.org/abs/1506.08594}{{\normalfont\ttfamily
  arXiv:1506.08594}}\relax
\mciteBstWouldAddEndPuncttrue
\mciteSetBstMidEndSepPunct{\mcitedefaultmidpunct}
{\mcitedefaultendpunct}{\mcitedefaultseppunct}\relax
\EndOfBibitem
\bibitem{Punzi:2004wh}
G.~Punzi, \ifthenelse{\boolean{articletitles}}{\emph{Comments on likelihood
  fits with variable resolution}, }{}eConf \textbf{C030908} (2003) WELT002,
  \href{http://arxiv.org/abs/physics/0401045}{{\normalfont\ttfamily
  arXiv:physics/0401045}}\relax
\mciteBstWouldAddEndPuncttrue
\mciteSetBstMidEndSepPunct{\mcitedefaultmidpunct}
{\mcitedefaultendpunct}{\mcitedefaultseppunct}\relax
\EndOfBibitem
\bibitem{LHCb-PAPER-2012-009}
LHCb collaboration, R.~Aaij {\em et~al.},
  \ifthenelse{\boolean{articletitles}}{\emph{{Measurement of the
  $D_s^+$--$D_s^-$ production asymmetry in 7 TeV $pp$ collisions}},
  }{}\href{http://dx.doi.org/10.1016/j.physletb.2012.06.001}{Phys.\ Lett.\
  \textbf{B713} (2012) 186},
  \href{http://arxiv.org/abs/1205.0897}{{\normalfont\ttfamily
  arXiv:1205.0897}}\relax
\mciteBstWouldAddEndPuncttrue
\mciteSetBstMidEndSepPunct{\mcitedefaultmidpunct}
{\mcitedefaultendpunct}{\mcitedefaultseppunct}\relax
\EndOfBibitem
\bibitem{cfit}
J.~Garra~Tico, \ifthenelse{\boolean{articletitles}}{\emph{The cfit fitting
  package}, }{} \url{http://www.github.com/cfit}\relax
\mciteBstWouldAddEndPuncttrue
\mciteSetBstMidEndSepPunct{\mcitedefaultmidpunct}
{\mcitedefaultendpunct}{\mcitedefaultseppunct}\relax
\EndOfBibitem
\bibitem{LHCb-PAPER-2012-031}
LHCb collaboration, {R.\ Aaij \textit{et al.\ \negmedspace\negmedspace}, and
  A.\ Bharucha} {\em et~al.},
  \ifthenelse{\boolean{articletitles}}{\emph{{Implications of LHCb measurements
  and future prospects}},
  }{}\href{http://dx.doi.org/10.1140/epjc/s10052-013-2373-2}{Eur.\ Phys.\ J.\
  \textbf{C73} (2013) 2373},
  \href{http://arxiv.org/abs/1208.3355}{{\normalfont\ttfamily
  arXiv:1208.3355}}\relax
\mciteBstWouldAddEndPuncttrue
\mciteSetBstMidEndSepPunct{\mcitedefaultmidpunct}
{\mcitedefaultendpunct}{\mcitedefaultseppunct}\relax
\EndOfBibitem
\end{mcitethebibliography}

\newpage


 
\newpage
\centerline{\large\bf LHCb collaboration}
\begin{flushleft}
\small
R.~Aaij$^{38}$, 
C.~Abell\'{a}n~Beteta$^{40}$, 
B.~Adeva$^{37}$, 
M.~Adinolfi$^{46}$, 
A.~Affolder$^{52}$, 
Z.~Ajaltouni$^{5}$, 
S.~Akar$^{6}$, 
J.~Albrecht$^{9}$, 
F.~Alessio$^{38}$, 
M.~Alexander$^{51}$
% ...

{\footnotesize \it
% ...
$ ^{5}$Clermont Universit\'{e}, Universit\'{e} Blaise Pascal, CNRS/IN2P3, LPC, Clermont-Ferrand, France\\
$ ^{6}$CPPM, Aix-Marseille Universit\'{e}, CNRS/IN2P3, Marseille, France\\
% ...
$ ^{9}$Fakult\"{a}t Physik, Technische Universit\"{a}t Dortmund, Dortmund, Germany\\
% ...
$ ^{37}$Universidad de Santiago de Compostela, Santiago de Compostela, Spain\\
$ ^{38}$European Organization for Nuclear Research (CERN), Geneva, Switzerland\\
% ...
$ ^{40}$Physik-Institut, Universit\"{a}t Z\"{u}rich, Z\"{u}rich, Switzerland\\
% ...
$ ^{46}$H.H. Wills Physics Laboratory, University of Bristol, Bristol, United Kingdom\\
% ...
$ ^{51}$School of Physics and Astronomy, University of Glasgow, Glasgow, United Kingdom\\
$ ^{52}$Oliver Lodge Laboratory, University of Liverpool, Liverpool, United Kingdom\\
% ...
}
\end{flushleft}

%
%

\end{document}